\documentclass{article}

\PassOptionsToPackage{numbers}{natbib}

\usepackage[preprint]{neurips_2022}

\usepackage{amsmath,amsfonts,bm}

\def\eqref#1{equation~\ref{#1}}

\def\1{\bm{1}}

\DeclareMathAlphabet{\mathsfit}{\encodingdefault}{\sfdefault}{m}{sl}
\SetMathAlphabet{\mathsfit}{bold}{\encodingdefault}{\sfdefault}{bx}{n}

\usepackage{hyperref}
\usepackage{url}
\usepackage{bbm}
\usepackage{graphicx}
\usepackage{soul}
\usepackage{amsmath}
\usepackage{amssymb}
\usepackage{microtype}
\usepackage{wrapfig}
\usepackage{xcolor}
\usepackage{booktabs}
\usepackage{multirow}
\usepackage{enumitem}

\usepackage{xspace}
\newcommand{\angstrom}{\r{A}\xspace}

\title{Contrastive Representation Learning\\for 3D Protein Structures}

\author{Pedro Hermosilla \& Timo Ropinski\\
Ulm University\\
}

\begin{document}

\maketitle

\begin{abstract}
Learning from 3D protein structures has gained wide interest in protein modeling and structural bioinformatics. 
Unfortunately, the number of available structures is orders of magnitude lower than the training data sizes commonly used in computer vision and machine learning. 
Moreover, this number is reduced even further, when only annotated protein structures can be considered, making the training of existing models difficult and prone to over-fitting. 
To address this challenge, we introduce a new representation learning framework for 3D protein structures. Our framework uses unsupervised contrastive learning to learn meaningful representations of protein structures, making use of proteins from the Protein Data Bank. 
We show, how these representations can be used to solve a large variety of tasks, such as protein function prediction, protein fold classification, structural similarity prediction, and protein-ligand binding affinity prediction. 
Moreover, we show how fine-tuned networks, pre-trained with our algorithm, lead to significantly improved task performance, achieving new state-of-the-art results in many tasks.
\end{abstract}

\section{Introduction}
In recent years, learning on 3D protein structures has gained a lot of attention in the fields of protein modeling and structural bioinformatics.
These neural network architectures process 3D positions of atoms and/or amino acids in order to make predictions of unprecedented performance, in tasks ranging from protein design~\citep{Ingraham2019genprot, strokach2020protdes, jing2021learning}, over protein structure classification~\citep{hermosilla2021ieconv}, protein quality assessment~\citep{baldassarre2020graphqa,derevyanko2018prot3dcnn}, and protein function prediction~\citep{amidi2017enzynet, gligorijevic2021function} -- just to name a few. 
Unfortunately, when learning on the structure of proteins one can only rely one a reduced amount of training data, as compared for example to sequence learning, since 3D structures are harder to obtain and thus less prevalent. 
While the Protein Data Bank (PDB)~\citep{berman2000pdb} today contains only around $182\,$K macromolecular structures, the Pfam database~\citep{mistry2020pfam} contains $47\,$M protein sequences. 
Naturally, the number of available structures decreases even further when only the structures labeled with a specific property are considered.
We refer to these as annotated protein structures. 
The SIFTS database, for example, contains around $220\,$K annotated enzymes from $96\,$K different PDB entries, and the SCOPe database contains 226\,K annotated structures. 
These numbers are orders of magnitude lower than the data set sizes which led to the major breakthroughs in the field of deep learning. ImageNet~\citep{ILSVRC15}, for instance, contains more than $10\,$M annotated images. 
As learning on 3D protein structures cannot benefit from these large amounts of data, model sizes are limited or over-fitting might occur, which can be avoided by making use of unlabeled data.

In order, to take advantage of unlabeled data, researchers have, over the years, designed different algorithms, that are able to learn meaningful representations from such data without labels~\citep{hadsell2006contrast, ye2019embedding, chen2020simple}. In natural language processing, next token prediction or random token masking are commonly used unsupervised training objectives, that are able to learn meaningful word representations useful for different downstream tasks~\citep{peters2018wordrep, devlin2019bert}. Recently, such algorithms have been used to learn meaningful protein representations from unlabeled sequences~\citep{alley2019unirep}, or as a pre-trained method for later fine-tuning models on different downstream tasks~\citep{rao2019tape}.
In computer vision recently, contrastive learning has shown great performance on image classification when used to pre-train deep convolutional neural network (CNN) architectures~\citep{chen2020simple,chen2020big}.
This pre-training objective has also been used in the context of protein sequence representation learning by dividing sequences in amino acid 'patches'~\citep{lu2020contrastiveprota}, or by using data augmentation techniques based on protein evolutionary information~\citep{lu2020contrastiveprotb}.
Most recently, the contrastive learning framework has been applied to graph convolutional neural networks~\citep{You2020GraphCL}.
These techniques were tested on protein spatial neighboring graphs (graphs where edges connect neighbor amino acids in 3D space) for the binary task of classifying a protein as an enzyme or not. 
However, these algorithms were designed for arbitrary graphs and did not take into account the underlying structure of proteins.

In this work, we introduce a contrastive learning framework for representation learning of 3D protein structures.
For each unlabeled protein chain, we select random molecular sub-structures during training.
We then minimize the cosine distance between the learned representations of the sub-structures sampled from the same protein, while maximizing the cosine distance between representations from different protein chains.
This training objective enables us, to pre-train models on all available annotated, but more importantly also unlabeled, protein structures. 
As we will show, the obtained representations can be used to improve performance on downstream tasks, such as structure classification, protein function, protein similarity, and protein-binding affinity prediction. 

The remainder of this paper is structured as follows.
First, we provide a summary of the state-of-the-art in Section~\ref{sec:related-work}.
Then, we introduce our framework in Section~\ref{sec:constrative-learning}.
Later, in Section~\ref{sec:experiments}, we describe the experiments conducted to evaluate our framework and the representations learned, and lastly, we provide a summary of our findings and possible lines of future research in Section~\ref{sec:conclusions}.

\section{Related Work}
\label{sec:related-work}

\noindent \textbf{3D protein structure learning.} Early work on learning from 3D protein structures used graph kernels and support vector machines to classify enzymes~\citep{borgwardt2005proteinds}.
Later, the advances in the fields of machine learning and computer vision brought a new set of techniques to the field.
Several authors represent the protein tertiary structure as a 3D density map, and process it with a 3D convolutional neural network (3DCNN).
Among the problems addressed with this approach, are protein quality assessment~\citep{derevyanko2018prot3dcnn}, protein enzyme classification~\citep{amidi2017enzynet}, protein-ligand binding affinity~\citep{ragoza2017prot3dcnn}, protein binding site prediction~\citep{jimenez2017deepsite} and protein-protein interaction interface prediction~\citep{townshend2019endtoend}.
Other authors have used graph convolutional neural networks (GCNN) to learn directly from the protein spatial neighboring graph.
Some of the tasks solved with these techniques, are protein interface prediction~\citep{fout2017interface}, function prediction~\citep{gligorijevic2021function}, protein quality assessment~\citep{baldassarre2020graphqa}, and protein design~\citep{strokach2020protdes}.
Recently, several neural network architectures, specifically designed for protein structures, have been proposed to tackle protein design challenges~\citep{Ingraham2019genprot, jing2021learning}, or protein fold and function prediction~\citep{hermosilla2021ieconv}.
However, all these methods have been trained with labeled data for downstream tasks.

\noindent \textbf{Protein representation learning.} Protein representation learning based on protein sequences is an active area of research.
Early works used similar techniques as the ones used in natural language processing to compute embeddings of groups of neighboring amino acids in a sequence~\citep{asgari2015bagofwords}.
Recently, other works have used unsupervised learning algorithms from natural language processing such as token masking or next token prediction~\citep{peters2018wordrep, devlin2019bert} to learn representations from protein sequences~\citep{alley2019unirep, rao2019tape, min2020pretrainstruct, strodthoff2020udsmprot}.
Recently, \citet{lu2020contrastiveprota, lu2020contrastiveprotb} have suggested using contrastive learning on protein sequences, to obtain a meaningful protein representation.
Despite the advances in representation learning for protein sequences, representation learning for 3D protein structures mostly has relied on hand-crafted features.
\citet{la20093dsurfer} proposed a method to compute a vector of 3D Zernike descriptors to represent protein surfaces, which later can be used for shape retrieval.
Moreover, \citet{guzenko2020realtime} used a similar approach, to compute a vector of 3D Zernike descriptors directly from the 3D density volume, which can be used later for protein shape comparison.
The annual shape retrieval contest (SHREC) usually contains a protein shape retrieval track, in which methods are required to determine protein similarity from different protein surfaces~\citep{biasotti2019proteinsshrec, biasotti2020proteinsshrec}.
Some of the works presented there, make use of 3DCNNs or GCNNs to achieve this goal.
However, they operate on protein surfaces, and are either trained in a supervised fashion, or pre-trained on a classification task.
Recently, \citet{Xia2022contrs} address the same problem by comparing protein graphs using contrastive learning based on protein similarity labels computed using TMAlign~\cite{zhang2005tm}.

To the best of our knowledge, the only works which attempted to use unsupervised pre-training algorithms in a structure-based model are the concurrent works of \citet{Wu2022md} and \citet{Zhang2022}. 
\citet{Wu2022md} propose to use molecular dynamics simulations on a small protein subset as a pre-training for the task of protein-ligand binding affinity prediction.
\citet{Zhang2022} focus on multiview contrast and self-prediction learning, whereby also the learned representations are not directly facilitated. In this paper, we will show how our framework is able to outperform these, even if the pre-training data set is filtered to remove similar proteins to the ones in the test sets, which also differs from concurrent work.

\noindent \textbf{Contrastive learning.} In 1992, \citet{becker1992contrast} suggested training neural networks through the agreement between representations of the same image under different transformations.
Later, \citet{hadsell2006contrast} proposed to learn image representations by minimizing the distance between positive pairs and maximizing the distance between negative pairs (see Figure~\ref{img:overview}).
This idea was used in other works by sampling negative pairs from the mini-batches used during training~\citep{ji2019invinf, ye2019embedding}.
Recently, \citet{chen2020simple, chen2020big} have shown how these methods can improve image classification performance. 
\citet{You2020GraphCL} have 
transferred these ideas to graphs, by proposing four different data transformations to be used during training: node dropping, edge perturbation, attribute masking, and subgraph sampling.
These ideas were tested on the commonly used graph benchmark \textsc{PROTEINS}~\citep{borgwardt2005proteinds}, composed of only $1,113$ proteins.
However, since this data set is composed of spatial neighboring graphs of secondary structures, the proposed data augmentation techniques can generate graphs of unconnected chain sections.
In this paper instead, we suggest using a domain-specific transformation strategy, that preserves the local information of protein sequences.

\section{3D Protein Contrastive Learning}
\label{sec:constrative-learning}

In this section, we describe our contrastive learning framework, composed of a domain-specific data augmentation algorithm used during pre-training and a neural network designed for protein structures.

\subsection{Protein graph}
\label{subsec:prot_graph}
In this work, the protein chain is defined as a graph $\mathcal G = (\mathcal N, \mathcal R, \mathcal F, \mathcal A, \mathcal B)$, where each node represents the alpha carbon of an amino acid with its 3D coordinates, $\mathcal N \in \mathbb R^{n \times 3}$, being $n$ the number of amino acids in the protein.
Moreover, for each node, we store a local frame composed of three orthonormal vectors describing the orientation of the amino acid wrt. the protein backbone, $\mathcal R \in \mathbb R^{n \times 3 \times 3}$.
Lastly, each node has also $t$ different associated features with it, $\mathcal F \in \mathbb R^{n \times t}$. 
The connectivity of the graph is stored in two different adjacency matrices, $\mathcal A \in \mathbb R^{n \times n}$ and $\mathcal B \in \mathbb R^{n \times n}$.
Matrix $\mathcal A$ is defined as $\mathcal A_{ij} = 1$ if amino acids $i$ and $j$ are connected by a peptide bond and $\mathcal A_{ij} = 0$ otherwise.
In matrix $\mathcal B$, $\mathcal B_{ij} = 1$ if amino acids $i$ and $j$ are at a distance smaller than $d$ in 3D space and $\mathcal B_{ij} = 0$ otherwise.

\subsection{Contrastive learning framework}
\label{sec:contrastive}

\begin{wrapfigure}{r}{0.5\textwidth}
\vspace{-.6cm}%
  \begin{center}
    \includegraphics[width=0.49\textwidth]{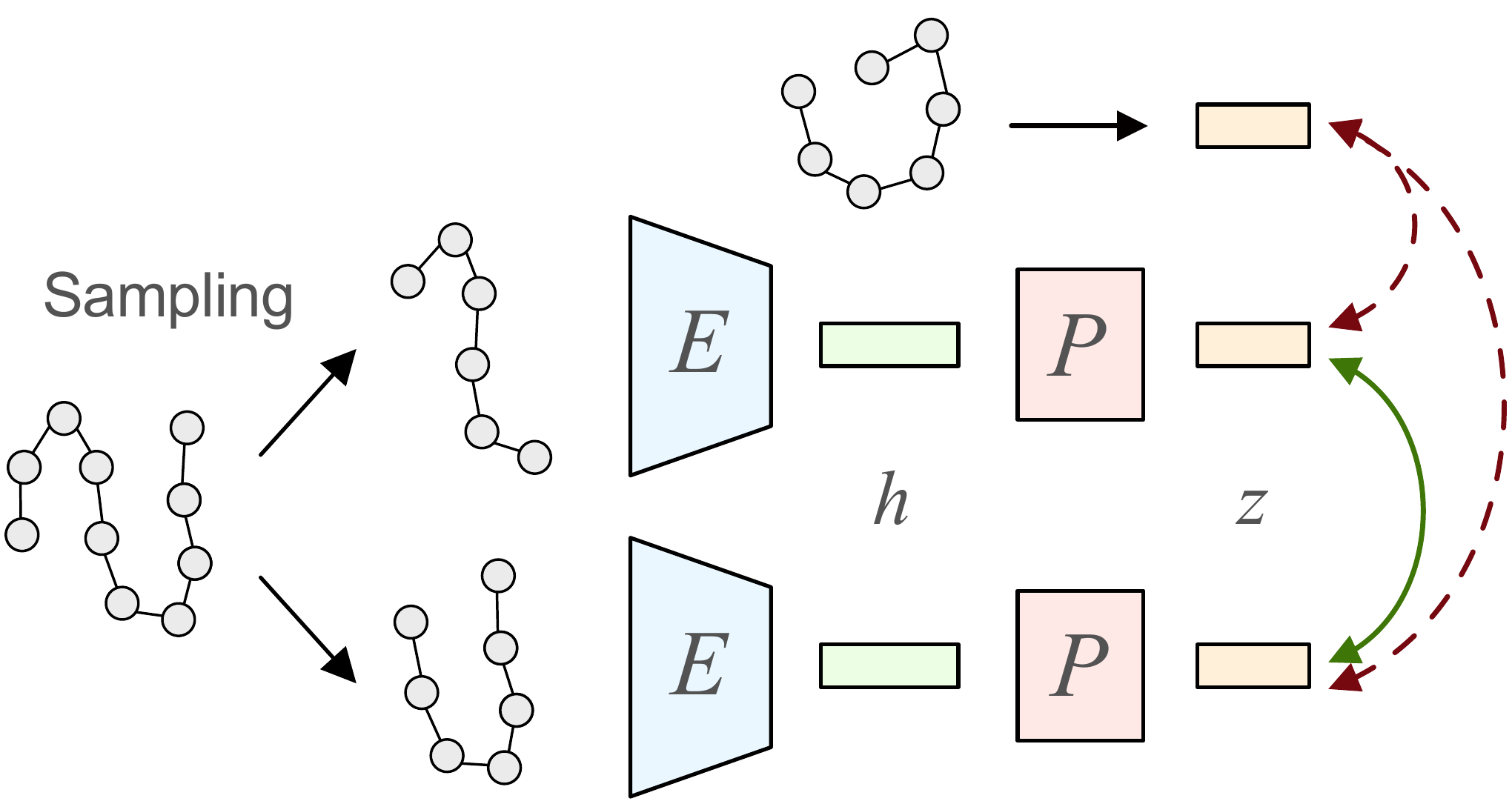}
      \vspace{-.2cm}%
      \caption{For each protein we sample random sub-structures which are then encoded into two representations, $h$ and $z$, using encoders $E$ and $P$. Then, we minimize the distance between representations $z$ from the same protein and maximize the distance between representations from different proteins.}
      \label{img:overview}
      \vspace{-.4cm}%
  \end{center}
\end{wrapfigure}

Inspired by recent works in computer vision~\citep{ye2019embedding, ji2019invinf, chen2020simple}, our framework is trained by maximizing the similarity between representations from sub-structures of the same protein, and minimizing the similarity between sub-structures from different proteins (see Fig.~\ref{img:overview}). 
More formally, given a protein graph $\mathcal G$, we sample two sub-structures $\mathcal G_i$ and $\mathcal G_j$ from it.
We then compute the latent representations of these sub-structures, $h_i$ and $h_j$, using a protein graph encoder, $h_i=E(\mathcal G_i)$.
Based on the findings of \citet{chen2020simple}, we further project these latent representations into smaller latent representation, $z_i$ and $z_j$, using a multilayer perceptron (MLP) with a single hidden layer, $z_i = P(h_i)$.
Lastly, the similarity between these representations is computed using the cosine distance, $s(z_i, z_j)$.
In order to minimize the similarity between these representations, we use the following loss function for the sub-structure $\mathcal G_i$:

\begin{equation}
    l_i = -log\frac{exp(s(z_i, z_j)/\tau)}{\sum_{k=1}^{2N}{\mathbbm{1}_{[k \neq i,k \neq j]} exp(s(z_i, z_k)/\tau)}}
    \label{eq:loss}
\end{equation}

\noindent where $\tau$ is a temperature parameter used to improve learning from 'hard' examples, $\mathbbm{1}_{[k \neq i,k \neq j]} \in [0,1]$ is a function that evaluates to $1$ if $k \neq i$ and $k \neq j$, and $N$ is the number of protein structures in the current mini-batch.
To compute $l_j$ we use again Equation~\ref{eq:loss}, but exchange the role of $i$ and $j$. 
This loss has been used before in the context of representation learning~\citep{chen2020simple, vandenoord2018replearn}, and, as in previous work, our framework does not explicitly sample negative examples but uses instead the rest of sub-structures sampled from different proteins in the mini-batch as negative examples.
In the following subsections, we will describe the different components specific to our framework designed to process protein structures.

\subsection{Sub-structure sampling}
\label{subsec:samplings}

As \citet{chen2020simple} demonstrated, the data transformation applied to the input, is of key importance to obtaining a meaningful representation.
In this work, we propose to use a domain-specific cropping strategy of the input data transformations.

\begin{wrapfigure}{r}{0.5\textwidth}
\vspace{-.6cm}%
  \begin{center}
    \includegraphics[width=0.48\textwidth]{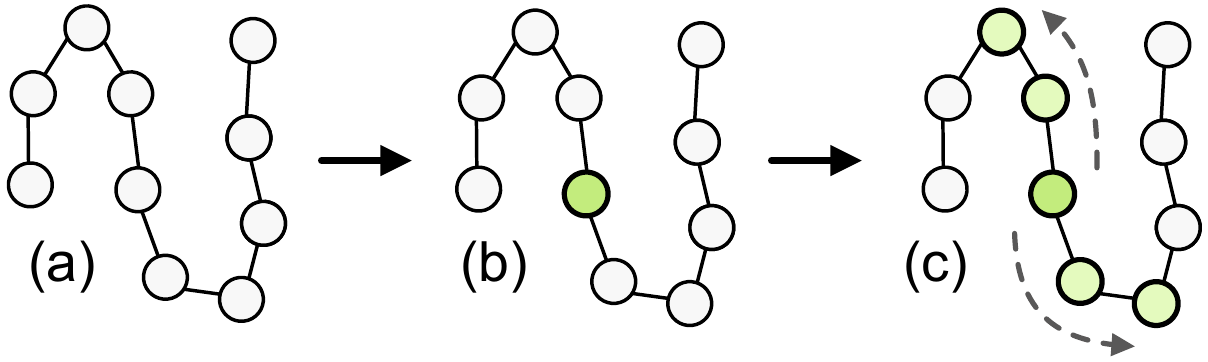}
      \vspace{-.2cm}%
      \caption{Our sampling strategy used during contrastive learning. For a protein chain \textbf{(a)}, we select a random amino acid \textbf{(b)}. Then we travel along the chain in both directions until we have a certain percentage $p$ of the sequence covered \textbf{(c)}.}
      \label{img:samplings}
      \vspace{-.4cm}%
  \end{center}
\end{wrapfigure}

Protein chains are composed of one or several stable sub-structures, called protein domains, which reoccur in different proteins.
These sub-structures can indicate evolutionary history between different proteins, as well as the function carried out by the protein~\citep{ponting2002domains}. 
Our sampling strategy uses the concept of protein sub-structures to sample for each protein two different continuous sub-structures along the polypeptide chain.
We achieve that, by first selecting a random amino acid in the protein chain $x_i \in \mathcal N$.
We then travel along the protein sequence in both directions using the adjacency matrix $\mathcal A$ while selecting each amino acid $x_{i+t}$ and $x_{i-t}$ in the process.
This process continues until we have covered a certain percentage $p$ of the protein chain, whereby our experiments indicate that a value of $p$ between $40\%$ and $60\%$ provides the best results (see supplementary material).
If during this sampling we reach the end of the sequence in one of the directions, we continue sampling in the other direction until we have covered the desired percentage $p$.
The selected amino acids compose the sub-structure that is then given as input to the graph encoder $\mathcal E$.
Figure ~\ref{img:samplings} illustrates this process.
Note that, since our framework learns from unlabeled data, we do not sample specifically protein domains from the protein chain, which would require annotations.
We instead sample random sub-structures that might be composed of a complete or partial domain, or, in large proteins, even span several domains.
The training objective then enforces a similar representation for random sub-structures of the same protein chain, where the properties of the complete structure have to be inferred.
We will show in our experiments, that these representations are able to encode features describing structural patterns and protein functions.

\subsection{Protein Encoder}
\label{subsec:encoder}

The information captured by a learned representation using contrastive learning strongly depends on the network architecture used to encode the input~\citep{tschannen2020contrast}.
Therefore, we design our protein encoder with properties that made neural networks successful in other fields, hierarchical feature computation~\cite{zeiler2014cnn} and transformation invariant/equivariant operations~\cite{deng2021vn}.
In the following paragraphs, we describe the proposed network architecture and operations.

\noindent \textbf{Network architecture.} Our protein encoder receives as input the protein graphs described in Sec.~\ref{subsec:prot_graph}.
First, we use an amino acid embedding, which is learned together with the parameters of the network, as the initial features.
These features are then processed by a set of  ResNet blocks~\citep{he2016resnet}.
Moreover, we use pooling operations to reduce the dimensionality of the graph as done in \citet{hermosilla2021ieconv}.
Two consecutive amino acids along the chain are pooled together into a single node in the pooled graph by averaging their 3D positions and features.
This process is repeated four times where the graph is pooled and the number of features per node increased.
More details on the network architecture are provided in the supplementary material.

\noindent \textbf{Convolution operation. }
Similar to \citet{hermosilla2021ieconv}, we define our convolution operation on the protein graph as follows.
For node $x_i$ on the graph, features of layer $l$ are computed by aggregating all features from the previous layer $l-1$ from all nodes of the graph $x_j$ at distance smaller than $d$ in 3D space from $x_i$.
Before aggregation, features from node $x_j$ are modulated by a kernel $k_o$ represented as a single layer MLP that takes as input the edge information between $x_i$ and $x_j$.
More formally:
\begin{equation}
    F_o^l (\mathcal G, x_i) = \sum_{j \in \mathcal N (x_i)} F_{j}^{l-1}k_o(f(\mathcal G, x_i, x_j))
    \label{eq:conv}
\end{equation}
where $\mathcal N (x_i)$ is the list of nodes at a distance smaller than $d$ from $x_i$, i.e. neighbor nodes in graph $\mathcal B$, and $f(\mathcal G, x_i, x_j)$ is the function that computes the edge information between node $x_i$ and $x_j$.

\noindent \textbf{Edge features. }
Function $f$, similar to \citet{Ingraham2019genprot}, computes the following edge information:
\begin{itemize}[leftmargin=.15in]
    \vspace{-0,1cm}
    \setlength\itemsep{0.05em}
    \item $\Vec{t}$:  The vector $x_j - x_i$ represented in the local frame of node $x_i$, $O_i \in \mathcal R$, and normalized by $d$.
    \item $r$: Dot product between the three axes of the local frames $O_i$ and $O_j$.
    \item $\delta$: The shortest path along peptide bond  between nodes $x_i$ and $x_j$, normalized by $\delta_{max}$.
    \vspace{-0,1cm}
\end{itemize}
These features are able to efficiently describe the relative position and orientation of neighboring node $x_j$ wrt. $x_i$, being translation invariant and rotation equivariant at the same time.

\noindent \textbf{Edge feature augmentation. }
The seven edge features computed by $f$ ($\Vec{t}$, $r$, and $\delta$) all have values in the range $[-1,1]$.
Similar to positional encoding~\citep{vaswani2017transf}, we further augment these inputs by applying the function $g = 1-2\lvert x \rvert$, which makes all features contribute to the final value of the kernel $k_o$ even when their values are equal to zero. 
This feature augmentation results in $14$ final input values to $k_o$, the original $7$ edges features, plus the transformed ones.

\noindent \textbf{Smooth receptive field. }
We weigh the resulting value of kernel $k_o$ by a function $\alpha$ to remove discontinuities at distance $d$, where a small displacement of a neighboring node $x_j$ can make it exit the receptive field.
Similar to the cutoff function proposed by \citet{klicpera2020direcmessage}, the function $\alpha$ smoothly decreases from one to zero at the edge of the receptive field, making the contributions of neighboring nodes disappear as they approach $d$.
Our function is defined as $\alpha = (1 - tanh(d_i*16 - 14))/2$, where $d_i$ is the distance of the neighboring node normalized by $d$.

\section{Experiments}
\label{sec:experiments}
In this section, we will describe the experiments conducted to evaluate our method, and demonstrate the value of the learned representations.
Our main data set used for unsupervised learning is acquired from the PDB~\cite{berman2000pdb}. 
We collected $476,362$ different protein chains, each composed of at least $25$ of amino acids. 
This set of protein chains was later reduced for each downstream task to avoid similarities with the different test sets, removing chains from the pre-training set based on the available annotations. 
For all downstream tasks, we measure the performance on three variants of our framework: using the protein encoder trained from scratch (\textit{no pre-train}), fixing the pre-trained protein encoder and learning a transformation of the representation with a multi-layer perceptron (\textit{MLP}), as well as fine-tuning the pre-trained protein encoder (\textit{fine-tune}).
For a detailed description of the experiments we refer the reader to the supplementary material.

\subsection{Protein structural similarity}

Protein similarity metrics are key in the study of the relationship between protein structure and function, and protein evolution. 
Predicting accurate protein similarities is an indication that a learned representation contains an accurate abstract representation of the 3D structure of the protein. 
Therefore, to validate our framework, we first use the pre-trained models on the downstream task of protein similarity prediction.
To this end, we use two different data sets, the DaliLite data set~\cite{holm2019dalilite} and the GraSR data set~\cite{Xia2022contrs}.
We use the same network architecture and setup as in our contrastive learning framework (Sec.~\ref{sec:contrastive}), where each protein pair is processed by our protein encoder and the similarity metric is defined as the cosine distance between the latent representations.

\begin{table}%
\vspace{-0,5cm}
\caption{Results of our method on the two protein structural similarity tasks. \textbf{Left:} Mean hit ratio of the first $1$ and $10$ proteins of each target for different learned distance metrics on the GraSR data set~\cite{Xia2022contrs}. \textbf{Right:} $F_{max}$ of different distance metrics with respect to Fold, Superfamily, and Family classifications on the DaliLite data set~\cite{holm2019dalilite}.}%
\setlength{\tabcolsep}{7pt}
\begin{center}
\begin{tabular}{lrr}
    \toprule
    &
    \multicolumn{1}{c}{1-HR} & 
    \multicolumn{1}{c}{10-HR}
    \\
    \midrule
    SGM~\cite{rogen2003gauss} & 0.275 & 0.285\\
    SSEF~\cite{zotenko2006secstruc} & 0.047 & 0.046\\
    DeepFold~\cite{Liu2018prot_struc_search} & 0.382 & 0.392\\
    GraSR~\cite{Xia2022contrs} & \underline{0.456} & 0.476\\
    \midrule
    Ours (\textit{no pre-train}) & 0.410 & 0.463\\
    Ours (\textit{MLP}) & 0.385 & \underline{0.480}\\
    Ours (\textit{fine-tune}) & \textbf{0.466} & \textbf{0.522}\\
    \bottomrule
\end{tabular}%
\hfill
\begin{tabular}{lrrr}
    \toprule
    &
    \multicolumn{1}{c}{Fold} & 
    \multicolumn{1}{c}{Super.} & 
    \multicolumn{1}{c}{Fam.}
    \\
    \midrule
    DaliLite~\cite{holm2019dalilite} & 
    0.38 & \textbf{0.83} & \underline{0.96}\\
    DeepAlign~\cite{wang2013protein} &
    0.28 & \underline{0.78} & \textbf{0.97}\\
    mTMaLign~\cite{dong2018mtm} &
    0.13 & 0.55 & 0.91\\
    TMaLign~\cite{zhang2005tm} & 
    0.12 & 0.39 & 0.85\\
    \midrule
    Ours (\textit{no pre-train}) & 
        \underline{0.63} & 0.62 & 0.66\\
    Ours (\textit{MLP}) & 
        \textbf{0.66} & 0.70 & 0.75\\
    Ours (\textit{fine-tune}) & 
        0.60 & 0.62 & 0.64\\
    \bottomrule
\end{tabular}%
\label{tbl:protein_sim}%
\end{center}
\end{table}

\paragraph{DaliLit dataset~\cite{holm2019dalilite}.} Here, for a given target protein in the test set, the model has to rank all proteins in the train data set based on their structural similarity.
The task measures how well the similarity metric captures the SCOPe classification hierarchy~\citep{murzin1995scope}, measuring the $F_{max}$ at different hierarchy levels: Fold, Superfamily, and Family. 
During training, we define similar proteins as all proteins belonging to the same fold. 
Therefore, we increase the cosine distance between proteins from the same fold and decreased it if they are from different folds.

The thus obtained results are illustrated in Tbl.~\ref{tbl:protein_sim} (right).
We can observe that our architecture without pre-training (\textit{no pre-train}) is able to achieve high $F_{max}$.
However, our pre-trained representations (\textit{MLP}) achieve better performance at all classification levels.
Surprisingly, fine-tuning the protein encoder on this task leads to a degradation in performance (\textit{fine-tune}).
Moreover, Tbl.~\ref{tbl:protein_sim} shows that our similarity metric captures the fold classification structure much better than other methods.
For the Superfamily classification scheme, our similarity metric achieves higher $F_{max}$ than commonly used similarity metrics such as TMAlign~\cite{zhang2005tm} or mTMAlign~\cite{dong2018mtm}.
Lastly, our method is not able to outperform other methods when measuring similarity at family level.
These results indicate that our learned similarity metric could facilitate the study of rare proteins with no similar known proteins. 
Furthermore, it is worth noticing that our metric is able to perform predictions orders of magnitude faster than the other methods.
When comparing timings for a single target, our method only takes a few seconds, as it just performs the dot product between the representations, plus around four minutes for initializing the system by loading and encoding of the proteins in the training set. 
In contrast, DaliLite~\citep{holm2019dalilite} requires $15$ hours and TMAlign~\citep{zhang2005tm} a bit less than one hour, on a computer equipped with six cores.

\paragraph{GraSR dataset~\cite{Xia2022contrs}.} Here, for a given target protein in the test set, the model also has to rank all proteins in the training set based on their structural similarity.
This data set considers a hit when the TMScore~\cite{zhang2005tm} between the target protein and the query protein is higher than $0.9*m$, where $m$ is the maximum TMScore between the target protein and all the query proteins in the training set.
Performance is measured with the mean hit ratio of the $1$ and $10$ most similar query proteins in the training set, as defined in \citet{Xia2022contrs}.
Since the TMScore is not symmetric, we use two different MLPs to transform the protein representation with different parameters, one for the query and another for the target proteins.
During training, we maximize the cosine distance between a target and query proteins considered as a hit and minimize it otherwise.

Tbl.~\ref{tbl:protein_sim} (left) presents the results obtained in this task.
Using the representations learned during pre-training (\textit{MLP}) improves performance over training the models from scratch (\textit{no pre-train}) for the $10$ hit ratio, and the highest $1$ and $10$ hit ratios are obtained when fine-tuning the pre-trained protein encoders (\textit{fine-tune}).
When compared to other methods developed to solve the same task, we can see that our pre-trained models achieve significantly higher hit ratios.

\subsection{Fold classification}

\begin{figure}[b]
  \begin{center}
    \includegraphics[width=\textwidth]{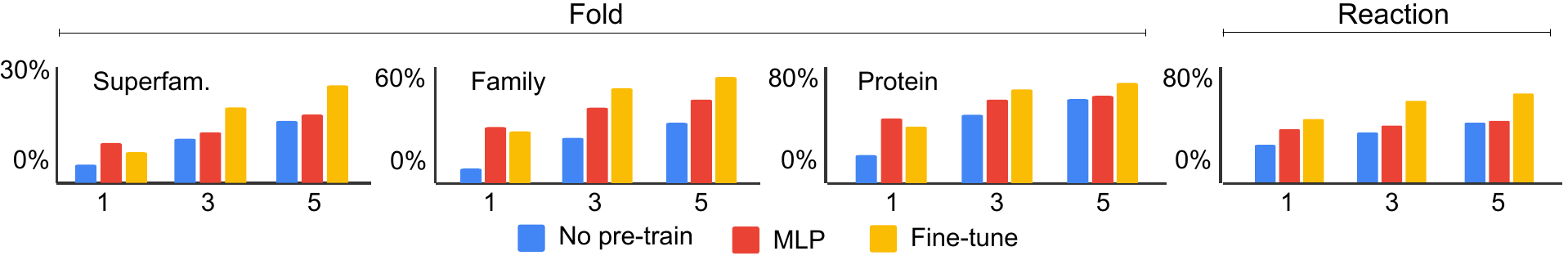}
      \caption{Accuracy on the Fold and Enzyme classification tasks wrt. the number of annotated proteins per class. Our pre-trained models improve generalization and reduce over-fitting on these cases.}
      \label{img:num_annot}
  \end{center}
\end{figure}

Protein fold classification and protein similarity metrics are key in structural bioinformatics to identify similar proteins. 
To evaluate our method on the protein fold classification task, we used the data set consolidated by \citet{hou2018deepsf} where the model has to predict the fold class of a protein among $1,195$ different folds. 
This data set contains three test sets with increasing difficulty based on the similarity between the proteins in the train and test sets:
Protein, Family, and Superfamily.
Performance is measured with mean accuracy on the test sets.
To solve this task, we use our proposed protein encoder to reduce the protein to a latent representation which is later used to classify the protein into a fold class.

Tbl.~\ref{tbl:fold_enzyme} presents the results obtained.
The results show that using the pre-trained representations (\textit{MLP}) for classification does not achieve higher accuracy than training the encoder from scratch (\textit{no pre-train}).
However, when the pre-trained model is fine-tuned, we achieve significantly higher accuracy (\textit{fine-tune}).
We can also see that our encoder achieves higher accuracy than state-of-the-art methods when trained from scratch or fine-tuned.
Lastly, we evaluate the robustness of the three versions of our framework when the number of annotated proteins per class is limited.
Fig.~\ref{img:num_annot} presents the results when only $1$, $3$, or $5$ proteins for each class are available during training.
Using the pre-trained representations (\textit{MLP}) achieves higher accuracy than training the model from scratch (\textit{no pre-train}) and higher accuracy than fine-tuning (\textit{fine-tune}) when only $1$ protein per class is available.
However, if the number of proteins is increased to $3$ or $5$, fine-tuning outperforms both.
These experiments illustrate that our pre-training framework improves generalization and reduces over-fitting when dealing with small data sets.

\begin{table}%
\parbox{.52\linewidth}{
\caption{Mean accuracy of our pre-trained networks on the Fold~\cite{hou2018deepsf} and Enzyme Reaction~\cite{hermosilla2021ieconv} classification tasks compared to other methods.}%
\setlength{\tabcolsep}{3pt}
\begin{center}
\begin{tabular}{lrrrr}
    \toprule
    &
    \multicolumn{3}{c}{\textsc{Fold}}&
    \multicolumn{1}{c}{\textsc{React.}}
    \\
    & 
    \multicolumn{1}{c}{Super.} & 
    \multicolumn{1}{c}{Fam.} & 
    \multicolumn{1}{c}{Prot.}&
    \\
    \midrule
    GCNN~\cite{kipf2016semi} &  
    16.8 & 21.3 & 82.8 &  67.3\\
    3DCNN~\cite{derevyanko2018prot3dcnn} &  
    31.6 & 45.4 & 92.5 &  78.8\\
    IEConv~\cite{hermosilla2021ieconv} & 
    45.0 & 69.7 & 98.9 &  \underline{87.2}\\
    \midrule
    Ours (\textit{no pre-train}) & 
    \underline{47.6} & \underline{70.2} & \underline{99.2} & \underline{87.2}\\
    Ours (\textit{MLP}) & 
    38.6 & 69.3 & 98.4 & 80.2\\
    Ours (\textit{fine-tune}) & 
    \textbf{50.3} &  \textbf{80.6} & \textbf{99.7} & \textbf{88.1}\\
    \bottomrule
\end{tabular}%
\label{tbl:fold_enzyme}%
\end{center}
}
\hfill
\parbox{.43\linewidth}{
\caption{$F_{max}$ of our method on the GO term prediction tasks~\cite{gligorijevic2021function} compared to other pre-training methods, with 3D structure ($^*$) or sequence information only ($^\dag$).}%
\setlength{\tabcolsep}{3pt}
\begin{center}
\begin{tabular}{lrrr}
    \toprule
    & 
    \multicolumn{1}{c}{MF} & 
    \multicolumn{1}{c}{BP} & 
    \multicolumn{1}{c}{CC}\\
    \midrule
    ESM-1b~\cite{rives2021esm}$^{\dag}$ & \underline{0.657} & \underline{0.470} & 0.488 \\
    LM-GVP~\cite{wang2022lmgvp}$^{\dag}$ & 0.545 & 0.417 & \textbf{0.527} \\
    GearNet~\cite{Zhang2022}$^{*}$ & 0.650 & \textbf{0.490} & 0.486 \\
    \midrule
    Ours (\textit{no pre-train}) & 0.624 & 0.421 & 0.431 \\
    Ours (\textit{MLP}) & 0.606 & 0.443 & 0.506 \\
    Ours (\textit{fine-tune}) & \textbf{0.661} & 0.468 & \underline{0.516}\\
    \bottomrule
\end{tabular}%
\label{tbl:go_terms}%
\end{center}
}
\end{table}%

\subsection{Protein function prediction}

Protein function prediction plays a crucial role in protein and drug design.
Being able to predict the function of a protein from its 3D structure directly allows determining functional information of \textit{de novo} proteins.
To accurately predict the functional information of proteins, the learned representations should contain fine-grained structural features describing such functions, making this task ideal to measure the expressiveness of the learned representations.
We evaluate our model on different data sets aimed to measure the prediction ability of models on different types of function annotations.
For all data sets, we use a protein encoder to reduce the protein into a latent representation which is later used to perform the final predictions.

\paragraph{Enzyme reaction~\cite{hermosilla2021ieconv}.} In this task, the model has to predict the reaction carried out by a protein enzyme among $384$ different classes, i.e., complete Enzyme Commission numbers (EC). 
The proteins in the data set are split into three sets, training, validation, and testing, whereby proteins in each set do not have more than $50\%$ of sequence similarity with proteins from the other sets. 
Performance is measured with mean accuracy on the test set. 

Tbl.~\ref{tbl:fold_enzyme} presents the results obtained for this task. 
We can see that fine-tuning our pre-trained model achieves the highest accuracy (\textit{fine-tune}), while using the pre-trained representations (\textit{MLP}) achieves competitive accuracy but does not outperform a model trained from scratch (\textit{no pre-train}).
Moreover, our framework achieves better accuracy than previous methods.
Lastly, we evaluated the performance when the number of annotated proteins per class is reduced to $1$, $3$, and $5$ (Fig.~\ref{img:num_annot}).
Our pre-trained models, fine-tuned or not, improve accuracy over a model trained from scratch in the three experiments.

\paragraph{GO terms~\cite{gligorijevic2021function}.}In this data set, the model has to determine the functions of a protein by predicting one or more Gene Ontology terms (GO).
This task is evaluated on three different data sets, where each one measures the performance on different types of GO terms, Molecular Function (MF), Biological Process (BP), and Cellular Component (CC).
The performance is measured with $F_{max}$.

Tbl.~\ref{tbl:go_terms} presents the results for the three data sets.
For the molecular function data set, our fine-tuned model achieves the higher $F_{max}$ (\textit{fine-tune}) while the pre-trained representations (\textit{MLP}) achieve similar performance as the model trained from scratch (\textit{no pre-train}).
For the biological process and cellular component data sets both pre-trained methods, fine-tuned and not, outperform models trained from scratch, being the fine-tuned model the one achieving the highest $F_{max}$.
When compared to other methods pre-trained on large sequence data sets of millions of proteins~\cite{rives2021esm,wang2022lmgvp}, our method outperforms those in the molecular function data set and achieves competitive performance on the other two. 
When compared to the pre-trained method of \citet{Zhang2022} on 3D structures, our framework outperforms it on two out of three data sets.
Lastly, we evaluate the performance of our models in relation to the sequence similarity between the train and the test sets.
While, in the molecular function data set we observe the same behavior at all levels of sequence similarity, in the biological process and cellular component data sets we observe that the difference in performance between pre-trained models and models trained from scratch increases as we decrease the sequence similarity, indicating that our pre-training algorithm improves generalization on proteins dissimilar to the train set.

\begin{figure}[t]
  \begin{center}
    \includegraphics[width=\textwidth]{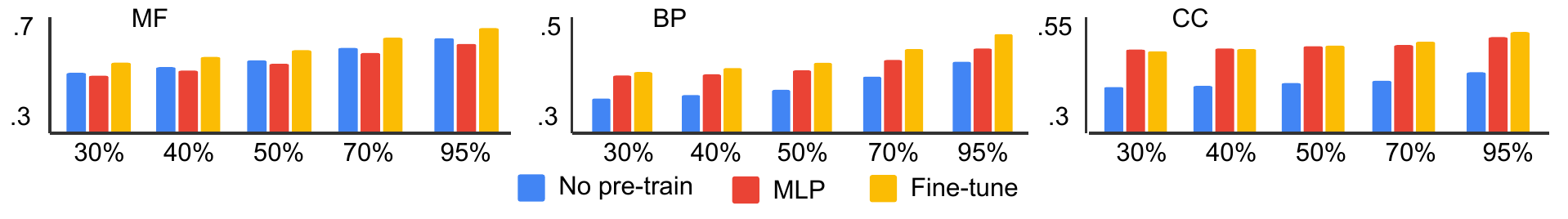}
      \caption{$F_{max}$ on the GO term prediction tasks wrt. the sequence similarity between the train and the test sets. Our pre-trained models improve generalization and reduce over-fitting on the Biological Process (BP) and the Cellular Component (CC) data sets when the sequence similarity decreases, while we do not observe a significant difference in the Molecular Function (MF) data set.}
      \label{img:seq_sim}
  \end{center}
\end{figure}

\subsection{Protein-Ligand binding affinity prediction}

Accurate prediction of the affinity between protein and ligands could accelerate the virtual screening of ligand candidates for drug discovery or the protein design process for proteins with specific functions.
To evaluate our model in this task we use the data set from \citet{townshend2022atom3d}.
In this task, the model has to predict the binding affinity between a protein and a ligand, expressed in molar units of the inhibition constant ($K_i$) or dissociation constant ($K_d$).
As in previous work~\cite{somnath2021multiscale,townshend2022atom3d,Wu2022md}, we do not distinguish between these constants and predict the negative log transformation of these, $pK =-log(K)$.
This task contains two different data sets with different maximum sequence similarity between the train and test set, $60\%$ and $30\%$.
Performance is measured with root mean squared error (RMSE), and Pearson and Spearman correlation coefficients.

To predict the binding affinity, we use the cosine distance between the learned representations of the protein and the ligand, scaled to the range defined by the maximum and minimum binding affinity in the data set.
The protein representation is obtained with the protein encoder described in Sec.~\ref{subsec:encoder} while the ligand representation is obtained with a three-layer graph neural network~\cite{kipf2016semi}.
Moreover, in order to avoid over-fitting in these small data sets, we reduce the number of layers to $3$ in the protein encoder.
For pre-training the ligand encoder, we use the \textit{in-vitro} subset of the ZINC20 database~\cite{irwin2020zinc20} with the same contrastive setup used for the protein encoder.

Tbl.~\ref{tbl:prot_lig} presents the results of these experiments.
We can see that using our pre-trained protein and ligand encoders, with and without fine-tuning, we achieve better results than using models trained from scratch, increasing generalization and reducing over-fitting.
Moreover, the fine-tuned models achieve the highest accuracy.
When compared to other methods, our models significantly improve the state-of-the-art on the $60\,\%$ sequence identity data set.
In the $30\,\%$ sequence identity data set, however, our method achieves competitive performance but it is not able to outperform other pre-trained models on molecular dynamics simulations~\cite{Wu2022md}.

\begin{table}%
\caption{RMSE, Pearson, and Spearman coeficients on the protein-ligand binding affinity prediction task~\cite{townshend2022atom3d}. Comparison to different 3D structure ($^*$) and sequence ($^\dag$)  based methods, and a method pretrained on molecular dynamics simulations (PretrainMD~\cite{Wu2022md}).}%
\setlength{\tabcolsep}{4.5pt}
\begin{center}
\footnotesize
\begin{tabular}{lrrrrrr}
    \toprule
    \toprule
    &
    \multicolumn{3}{c}{Seq. Id. ($60\,\%$)} &
    \multicolumn{3}{c}{Seq. Id. ($30\,\%$)}
    \\
    &
    \multicolumn{1}{c}{RMSE $\downarrow$} &
    \multicolumn{1}{c}{Pears. $\uparrow$} &
    \multicolumn{1}{c}{Spear. $\uparrow$} &
    \multicolumn{1}{c}{RMSE $\downarrow$} &
    \multicolumn{1}{c}{Pears. $\uparrow$} &
    \multicolumn{1}{c}{Spear. $\uparrow$}
    \\
    \midrule
    DeepDTA~\cite{ozturk2018deepdta}$^\dag$ & 
    1.762 {$\pm$ \scriptsize .261} & 0.666 {$\pm$ \scriptsize .012} & 0.663 {$\pm$ \scriptsize .015} &
     1.565 {$\pm$ \scriptsize .080} & \underline{0.573 {$\pm$ \scriptsize .022}} & 0.574 {$\pm$ \scriptsize .024}\\
    3DCNN~\cite{townshend2022atom3d}$^*$ & 
    1.450 {$\pm$ \scriptsize .024} & 0.716 {$\pm$ \scriptsize .008} & 0.714 {$\pm$ \scriptsize .009} &
    \underline{1.429 {$\pm$ \scriptsize .042}} & 0.541 {$\pm$ \scriptsize .029} & 0.532 {$\pm$ \scriptsize .033}\\
    3DGCNN~\cite{townshend2022atom3d}$^*$ & 
    1.493 {$\pm$ \scriptsize .010} & 0.669 {$\pm$ \scriptsize .013} & 0.691 {$\pm$ \scriptsize .010} &
    1.963 {$\pm$ \scriptsize .120} & \textbf{0.581 {$\pm$ \scriptsize .039}} & \textbf{0.647 {$\pm$ \scriptsize .071}}\\
    HoloProt~\cite{somnath2021multiscale}$^*$ & 
    1.365 {$\pm$ \scriptsize .038} & 0.749 {$\pm$ \scriptsize .014} & 0.742 {$\pm$ \scriptsize .011} &
    1.464 {$\pm$ \scriptsize .006} & 0.509 {$\pm$ \scriptsize .002} & 0.500 {$\pm$ \scriptsize .005}\\
    PretrainMD~\cite{Wu2022md}$^*$ & 
    1.468 {$\pm$ \scriptsize .026} & 0.673 {$\pm$ \scriptsize .015} & 0.691 {$\pm$ \scriptsize .014} &
    \textbf{1.419 {$\pm$ \scriptsize .027}} & 0.551 {$\pm$ \scriptsize .045} & \underline{0.575 {$\pm$ \scriptsize .033}}\\
    \midrule
    Ours (\textit{no pre-train}) & 
    \underline{1.347 {$\pm$ \scriptsize .018}} & 0.757 {$\pm$ \scriptsize .005} & 0.747 {$\pm$ \scriptsize .004} &
    1.589 {$\pm$ \scriptsize .081} & 0.455 {$\pm$ \scriptsize .045} & 0.451 {$\pm$ \scriptsize .043}\\
    Ours (\textit{MLP}) & 
    1.361 {$\pm$ \scriptsize .032} & \underline{0.763 {$\pm$ \scriptsize .009}} & \underline{0.763 {$\pm$ \scriptsize .010}} &
    1.525 {$\pm$ \scriptsize .070} & 0.498 {$\pm$ \scriptsize .036} & 0.493 {$\pm$ \scriptsize .044}\\
    Ours (\textit{fine-tune}) & 
    \textbf{1.332 {$\pm$ \scriptsize .020}} & \textbf{0.768 {$\pm$ \scriptsize .006}} & \textbf{0.764 {$\pm$ \scriptsize .006}} &
    1.452 {$\pm$ \scriptsize .044} & 0.545 {$\pm$ \scriptsize .023} & 0.532 {$\pm$ \scriptsize .025}\\
    \bottomrule
\end{tabular}%
\label{tbl:prot_lig}%
\end{center}
\end{table}

\section{Conclusions}
\label{sec:conclusions}
In this paper, we have introduced contrastive learning for protein structures. 
While learning on protein structures has shown remarkable results, it suffers from a rather low availability of annotated data sets, which increases the demand for unsupervised learning technologies. 
In this paper, we demonstrated, that by combining protein-aware data transformations with state-of-the-art learning technologies, we were able to obtain a learned representation without the need for such annotated data. 
This is highly beneficial, since the availability of annotated 3D structures is limited, as compared to sequence data. 
Moreover, we have shown that using our pre-trained models we can achieve new state-of-the-art performance on a large set of relevant protein tasks.

We believe that our work is a first important step in transferring unsupervised learning methods to large-scale protein structure databases. 
In the future, we foresee, that the learned representation can not only be used, to solve the tasks demonstrated in this paper, but that it can also be helpful, to solve other protein structure problems.
Protein-protein interaction prediction, for example, could be addressed using the cosine distance between the learned representations.
Additionally, upon acceptance, we plan to release the representations for all PDB proteins, and make our technology available, such that these representations can be updated with newly discovered proteins.

\newpage

\bibliography{main}

\begin{thebibliography}{72}
\providecommand{\natexlab}[1]{#1}
\providecommand{\url}[1]{\texttt{#1}}
\expandafter\ifx\csname urlstyle\endcsname\relax
  \providecommand{\doi}[1]{doi: #1}\else
  \providecommand{\doi}{doi: \begingroup \urlstyle{rm}\Url}\fi

\bibitem[Alley et~al.(2019)Alley, Khimulya, Biswas, AlQuraishi, and
  Church]{alley2019unirep}
Ethan~C. Alley, Grigory Khimulya, Surojit Biswas, Mohammed AlQuraishi, and
  George~M. Church.
\newblock Unified rational protein engineering with sequence-based deep
  representation learning.
\newblock \emph{Nature Methods}, 2019.

\bibitem[Amidi et~al.(2017)Amidi, Amidi, Vlachakis, Megalooikonomou, Paragios,
  and Zacharaki]{amidi2017enzynet}
A.~Amidi, S.~Amidi, D.~Vlachakis, V.~Megalooikonomou, N.~Paragios, and
  E.~Zacharaki.
\newblock {EnzyNet}: enzyme classification using {3D} convolutional neural
  networks on spatial representation.
\newblock \emph{arXiv:1707.06017}, 2017.

\bibitem[Asgari and Mofrad(2015)]{asgari2015bagofwords}
E.~Asgari and M.R.K. Mofrad.
\newblock Continuous distributed representation of biological sequences for
  deep proteomics and genomics.
\newblock \emph{PLoS ONE}, 2015.

\bibitem[Baldassarre et~al.(2020)Baldassarre, Hurtado, Elofsson, and
  Azizpour]{baldassarre2020graphqa}
F.~Baldassarre, D.~M. Hurtado, A.~Elofsson, and H.~Azizpour.
\newblock {GraphQA: Protein Model Quality Assessment using Graph Convolutional
  Networks}.
\newblock \emph{Bioinformatics}, 2020.

\bibitem[Becker and Hinton(1992)]{becker1992contrast}
Suzanna Becker and Geoffrey~E. Hinton.
\newblock Self-organizing neural network that discovers surfaces in random-dot
  stereograms.
\newblock \emph{Nature}, 1992.

\bibitem[Bepler and Berger(2019)]{bepler2019embedstruct}
Tristan Bepler and Bonnie Berger.
\newblock Learning protein sequence embeddings using information from
  structure.
\newblock \emph{International Conference on Learning Representations}, 2019.

\bibitem[Berman et~al.(2000)Berman, Westbrook, Feng, Gilliland, Bhat, Weissig,
  Shindyalov, and Bourne]{berman2000pdb}
Helen~M. Berman, John Westbrook, Zukang Feng, Gary Gilliland, T.~N. Bhat, Helge
  Weissig, Ilya~N. Shindyalov, and Philip~E. Bourne.
\newblock {The Protein Data Bank}.
\newblock \emph{Nucleic Acids Research}, 2000.

\bibitem[Borgwardt et~al.(2005)Borgwardt, Ong, Schönauer, Vishwanathan, Smola,
  and Kriegel]{borgwardt2005proteinds}
Karsten~M. Borgwardt, Cheng~Soon Ong, Stefan Schönauer, S.~V.~N. Vishwanathan,
  Alex~J. Smola, and Hans-Peter Kriegel.
\newblock {Protein function prediction via graph kernels}.
\newblock \emph{Bioinformatics}, 2005.

\bibitem[Chen et~al.(2020{\natexlab{a}})Chen, Kornblith, Norouzi, and
  Hinton]{chen2020simple}
Ting Chen, Simon Kornblith, Mohammad Norouzi, and Geoffrey Hinton.
\newblock A simple framework for contrastive learning of visual
  representations.
\newblock \emph{arXiv preprint arXiv:2002.05709}, 2020{\natexlab{a}}.

\bibitem[Chen et~al.(2020{\natexlab{b}})Chen, Kornblith, Swersky, Norouzi, and
  Hinton]{chen2020big}
Ting Chen, Simon Kornblith, Kevin Swersky, Mohammad Norouzi, and Geoffrey
  Hinton.
\newblock Big self-supervised models are strong semi-supervised learners.
\newblock \emph{arXiv preprint arXiv:2006.10029}, 2020{\natexlab{b}}.

\bibitem[Dana et~al.(2018)Dana, Gutmanas, Tyagi, Qi, O’Donovan, Martin, and
  Velankar]{Dana2019SIFTS}
J.~M Dana, A.~Gutmanas, N.~Tyagi, G.~Qi, C.~O’Donovan, M.~Martin, and
  S.~Velankar.
\newblock {SIFTS: updated Structure Integration with Function, Taxonomy and
  Sequences resource allows 40-fold increase in coverage of structure-based
  annotations for proteins}.
\newblock \emph{Nucleic Acids Research}, 2018.

\bibitem[Deng et~al.(2021)Deng, Litany, Duan, Poulenard, Tagliasacchi, and
  Guibas]{deng2021vn}
Congyue Deng, Or~Litany, Yueqi Duan, Adrien Poulenard, Andrea Tagliasacchi, and
  Leonidas Guibas.
\newblock Vector neurons: a general framework for so(3)-equivariant networks.
\newblock \emph{International Conference on Computer Vision (ICCV)}, 2021.

\bibitem[Derevyanko et~al.(2018)Derevyanko, Grudinin, Bengio, and
  Lamoureux]{derevyanko2018prot3dcnn}
Georgy Derevyanko, Sergei Grudinin, Yoshua Bengio, and Guillaume Lamoureux.
\newblock {Deep convolutional networks for quality assessment of protein
  folds}.
\newblock \emph{Bioinformatics}, 34\penalty0 (23):\penalty0 4046--53, 2018.

\bibitem[Devlin et~al.(2019)Devlin, Chang, Lee, and Toutanova]{devlin2019bert}
Jacob Devlin, Ming{-}Wei Chang, Kenton Lee, and Kristina Toutanova.
\newblock {BERT:} pre-training of deep bidirectional transformers for language
  understanding.
\newblock \emph{Annual Conference of the North American Chapter of the
  Association for Computational Linguistics (NAACL)}, 2019.

\bibitem[Diehl(2019)]{diehl2019edge}
Frederik Diehl.
\newblock Edge contraction pooling for graph neural networks.
\newblock \emph{arxiv:1905.10990}, 2019.

\bibitem[Dong et~al.(2018)Dong, Pan, Peng, Zhang, and Yang]{dong2018mtm}
Runze Dong, Shuo Pan, Zhenling Peng, Yang Zhang, and Jianyi Yang.
\newblock mtm-align: a server for fast protein structure database search and
  multiple protein structure alignment.
\newblock \emph{Nucleic acids research}, 2018.

\bibitem[Elnaggar et~al.(2020)Elnaggar, Heinzinger, Dallago, Rihawi, Wang,
  Jones, Gibbs, Feher, Angerer, Steinegger, Bhowmik, and
  Rost]{elnaggar2020prottrans}
A.~Elnaggar, M.~Heinzinger, C.~Dallago, G.~Rihawi, Y.~Wang, L.~Jones, T.~Gibbs,
  T.~Feher, C.~Angerer, M.~Steinegger, D.~Bhowmik, and B.~Rost.
\newblock Prottrans: Towards cracking the language of life's code through
  self-supervised deep learning and high performance computing.
\newblock 2020.

\bibitem[Fout et~al.(2017)Fout, Byrd, Shariat, and Ben-Hur]{fout2017interface}
A.~Fout, J.~Byrd, B.~Shariat, and A.~Ben-Hur.
\newblock Protein interface prediction using graph convolutional networks.
\newblock In \emph{Advances in Neural Information Processing Systems 30}. 2017.

\bibitem[Gainza et~al.(2020)Gainza, Sverrisson, Monti, Rodolà, Boscaini,
  Bronstein, and Correia]{gainza2020masif}
P.~Gainza, F.~Sverrisson, F.~Monti, E.~Rodolà, D.~Boscaini, M.~M. Bronstein,
  and B.~E. Correia.
\newblock Deciphering interaction fingerprints from protein molecular surfaces
  using geometric deep learning.
\newblock \emph{Nature Methods}, 2020.

\bibitem[Gligorijevic et~al.(2021)Gligorijevic, Renfrew, Kosciolek, Leman, Cho,
  Vatanen, Berenberg, Taylor, Fisk, Xavier, Knight, and
  Bonneau]{gligorijevic2021function}
V.~Gligorijevic, P.~D. Renfrew, T.~Kosciolek, J.~K. Leman, K.~Cho, T.~Vatanen,
  D.~Berenberg, B.~Taylor, I.~M. Fisk, R.~J. Xavier, R.~Knight, and R.~Bonneau.
\newblock Structure-based function prediction using graph convolutional
  networks.
\newblock \emph{Nature Communications}, 2021.

\bibitem[Guzenko et~al.(2020)Guzenko, Burley, and Duarte]{guzenko2020realtime}
Dmytro Guzenko, Stephen~K. Burley, and Jose~M. Duarte.
\newblock Real time structural search of the protein data bank.
\newblock \emph{PLOS Computational Biology}, 2020.

\bibitem[Hadsell et~al.(2006)Hadsell, Chopra, and LeCun]{hadsell2006contrast}
R.~Hadsell, S.~Chopra, and Y.~LeCun.
\newblock Dimensionality reduction by learning an invariant mapping.
\newblock In \emph{IEEE Computer Society Conference on Computer Vision and
  Pattern Recognition (CVPR)}, 2006.

\bibitem[He et~al.(2016)He, Zhang, Ren, and Sun]{he2016resnet}
Kaiming He, Xiangyu Zhang, Shaoqing Ren, and Jian Sun.
\newblock Deep residual learning for image recognition.
\newblock In \emph{2016 IEEE Conference on Computer Vision and Pattern
  Recognition (CVPR)}, 2016.

\bibitem[Hermosilla et~al.(2021)Hermosilla, Schäfer, Lang, Fackelmann,
  Vázquez, Kozlíková, Krone, Ritschel, and Ropinski]{hermosilla2021ieconv}
Pedro Hermosilla, Marco Schäfer, Matěj Lang, Gloria Fackelmann, Pere~Pau
  Vázquez, Barbora Kozlíková, Michael Krone, Tobias Ritschel, and Timo
  Ropinski.
\newblock Intrinsic-extrinsic convolution and pooling for learning on 3d
  protein structures.
\newblock \emph{International Conference on Learning Representations}, 2021.

\bibitem[Holm(2019)]{holm2019dalilite}
Liisa Holm.
\newblock {Benchmarking fold detection by DaliLite v.5}.
\newblock \emph{Bioinformatics}, 2019.

\bibitem[Hou et~al.(2018)Hou, Adhikari, and Cheng]{hou2018deepsf}
J.~Hou, B.~Adhikari, and J.~Cheng.
\newblock Deepsf: Deep convolutional neural network for mapping protein
  sequences to folds.
\newblock In \emph{Proceedings of the 2018 ACM International Conference on
  Bioinformatics, Computational Biology, and Health Informatics}, 2018.

\bibitem[Ingraham et~al.(2019)Ingraham, Garg, Barzilay, and
  Jaakkola]{Ingraham2019genprot}
John Ingraham, Vikas Garg, Regina Barzilay, and Tommi Jaakkola.
\newblock Generative models for graph-based protein design.
\newblock In \emph{NeurIPS}, pages 15820--15831, 2019.

\bibitem[Irwin et~al.(2020)Irwin, Tang, Young, Dandarchuluun, Wong,
  Khurelbaatar, Moroz, Mayfield, and Sayle]{irwin2020zinc20}
John~J. Irwin, Khanh~G. Tang, Jennifer Young, Chinzorig Dandarchuluun,
  Benjamin~R. Wong, Munkhzul Khurelbaatar, Yurii~S. Moroz, John Mayfield, and
  Roger~A. Sayle.
\newblock Zinc20—a free ultralarge-scale chemical database for ligand
  discovery.
\newblock \emph{Journal of Chemical Information and Modeling}, 2020.

\bibitem[Ji et~al.(2019)Ji, Henriques, and Vedaldi]{ji2019invinf}
Xu~Ji, Jo{\~{a}}o~F. Henriques, and Andrea Vedaldi.
\newblock Invariant information distillation for unsupervised image
  segmentation and clustering.
\newblock \emph{International Conference on Computer Vision (ICCV)}, 2019.

\bibitem[Jiménez et~al.(2017)Jiménez, Doerr, Martínez-Rosell, and
  Rose]{jimenez2017deepsite}
J~Jiménez, S~Doerr, G~Martínez-Rosell, and A~S Rose.
\newblock {DeepSite}: protein-binding site predictor using {3D}-convolutional
  neural networks.
\newblock \emph{Bioinformatics}, 2017.

\bibitem[Jing et~al.(2021)Jing, Eismann, Suriana, Townshend, and
  Dror]{jing2021learning}
Bowen Jing, Stephan Eismann, Patricia Suriana, Raphael John~Lamarre Townshend,
  and Ron Dror.
\newblock Learning from protein structure with geometric vector perceptrons.
\newblock In \emph{International Conference on Learning Representations}, 2021.

\bibitem[Kipf and Welling(2017)]{kipf2016semi}
Thomas~N Kipf and Max Welling.
\newblock Semi-supervised classification with graph convolutional networks.
\newblock \emph{ICML}, 2017.

\bibitem[Klicpera et~al.(2020)Klicpera, Gro{\ss}, and
  G{\"{u}}nnemann]{klicpera2020direcmessage}
Johannes Klicpera, Janek Gro{\ss}, and Stephan G{\"{u}}nnemann.
\newblock Directional message passing for molecular graphs.
\newblock \emph{Internation Conference on Learning Representations}, 2020.

\bibitem[La et~al.(2009)La, Esquivel-Rodríguez, Venkatraman, Li, Sael, Ueng,
  Ahrendt, and Kihara]{la20093dsurfer}
David La, Juan Esquivel-Rodríguez, Vishwesh Venkatraman, Bin Li, Lee Sael,
  Stephen Ueng, Steven Ahrendt, and Daisuke Kihara.
\newblock {3D-SURFER: software for high-throughput protein surface comparison
  and analysis}.
\newblock \emph{Bioinformatics}, 2009.

\bibitem[Langenfeld et~al.(2019)Langenfeld, Axenopoulos, Benhabiles, Daras,
  Giachetti, Han, Hammoudi, Kihara, Lai, Liu, Melkemi, Mylonas, Terashi, Wang,
  Windal, and Montes]{biasotti2019proteinsshrec}
Florent Langenfeld, Apostolos Axenopoulos, Halim Benhabiles, Petros Daras,
  Andrea Giachetti, Xusi Han, Karim Hammoudi, Daisuke Kihara, Tuan~M. Lai,
  Haiguang Liu, Mahmoud Melkemi, Stelios~K. Mylonas, Genki Terashi, Yufan Wang,
  Feryal Windal, and Matthieu Montes.
\newblock {Protein Shape Retrieval Contest}.
\newblock In \emph{Eurographics Workshop on 3D Object Retrieval}, 2019.

\bibitem[Langenfeld et~al.(2020)Langenfeld, Peng, Lai, Rosin, Aderinwale,
  Terashi, Christoffer, Kihara, Benhabiles, Hammoudi, Cabani, Windal, Melkemi,
  Giachetti, Mylonas, Axenopoulos, Daras, Otu, Zwiggelaar, and
  Hunter]{biasotti2020proteinsshrec}
Florent Langenfeld, Yuxu Peng, Yu~Kun Lai, Paul~L. Rosin, Tunde Aderinwale,
  Genki Terashi, Charles Christoffer, Daisuke Kihara, Halim Benhabiles, Karim
  Hammoudi, Adnane Cabani, Feryal Windal, Mahmoud Melkemi, Andrea Giachetti,
  Stelios Mylonas, Apostolos Axenopoulos, Petros Daras, Ekpo Otu, Reyer
  Zwiggelaar, and David Hunter.
\newblock {Multi-domain protein shape retrieval challenge}.
\newblock In Silvia Biasotti, Guillaume Lavoué, and Remco Veltkamp, editors,
  \emph{Eurographics Workshop on 3D Object Retrieval}, 2020.

\bibitem[Liu et~al.(2018)Liu, Ye, Wang, and Peng]{Liu2018prot_struc_search}
Yang Liu, Qing Ye, Liwei Wang, and Jian Peng.
\newblock {Learning structural motif representations for efficient protein
  structure search}.
\newblock \emph{Bioinformatics}, 2018.

\bibitem[Liu et~al.(2017)Liu, Su, Han, Liu, Yang, Li, and Wang]{liu2017pdbbind}
Zhihai Liu, Minyi Su, Li~Han, Jie Liu, Qifan Yang, Yan Li, and Renxiao Wang.
\newblock Forging the basis for developing protein–ligand interaction scoring
  functions.
\newblock \emph{Accounts of Chemical Research}, 2017.

\bibitem[Lu et~al.(2020{\natexlab{a}})Lu, Lu, and
  Moses]{lu2020contrastiveprotb}
A.~X. Lu, A.~X. Lu, and A.~Moses.
\newblock Evolution is all you need: Phylogenetic augmentation for contrastive
  learning.
\newblock \emph{15th Machine Learning in Computational Biology (MLCB)},
  2020{\natexlab{a}}.

\bibitem[Lu et~al.(2020{\natexlab{b}})Lu, Zhang, Ghassemi, and
  Moses]{lu2020contrastiveprota}
A.~X. Lu, H.~Zhang, M.~Ghassemi, and A.~Moses.
\newblock A self-supervised contrastive learning of protein representations by
  mutual information maximization.
\newblock \emph{15th Machine Learning in Computational Biology (MLCB)},
  2020{\natexlab{b}}.

\bibitem[Min et~al.(2020)Min, Park, Kim, Choi, and Yoon]{min2020pretrainstruct}
Seonwoo Min, Seunghyun Park, Siwon Kim, Hyun-Soo Choi, and Sungroh Yoon.
\newblock Pre-training of deep bidirectional protein sequence representations
  with structural information.
\newblock \emph{Bioinformatics}, 2020.

\bibitem[Mistry et~al.(2020)Mistry, Chuguransky, Williams, Qureshi, Salazar,
  Sonnhammer, Tosatto, Paladin, Raj, Richardson, Finn, and
  Bateman]{mistry2020pfam}
Jaina Mistry, Sara Chuguransky, Lowri Williams, Matloob Qureshi, Gustavo A
  Salazar, Erik L~L Sonnhammer, Silvio C~E Tosatto, Lisanna Paladin, Shriya
  Raj, Lorna~J Richardson, Robert~D Finn, and Alex Bateman.
\newblock {Pfam: The protein families database in 2021}.
\newblock \emph{Nucleic Acids Research}, 2020.

\bibitem[Murzin et~al.(1955)Murzin, Brenner, Hubbard, and
  Chothia]{murzin1995scope}
A.G. Murzin, S.E. Brenner, T.~Hubbard, and C.~Chothia.
\newblock {SCOP}: a structural classification of proteins database for the
  investigation of sequences and structures.
\newblock \emph{J Molecular Biology}, 1955.

\bibitem[Peters et~al.(2018)Peters, Neumann, Iyyer, Gardner, Clark, Lee, and
  Zettlemoyer]{peters2018wordrep}
Matthew~E. Peters, Mark Neumann, Mohit Iyyer, Matt Gardner, Christopher Clark,
  Kenton Lee, and Luke Zettlemoyer.
\newblock Deep contextualized word representations.
\newblock \emph{Annual Conference of the North American Chapter of the
  Association for Computational Linguistics (NAACL)}, 2018.

\bibitem[Ponting and Russell(2002)]{ponting2002domains}
Chris~P. Ponting and Robert~R. Russell.
\newblock The natural history of protein domains.
\newblock \emph{Annual Review of Biophysics and Biomolecular Structure}, 2002.

\bibitem[Ragoza et~al.(2017)Ragoza, Hochuli, Idrobo, Sunseri, and
  Koes]{ragoza2017prot3dcnn}
Matthew Ragoza, Joshua Hochuli, Elisa Idrobo, Jocelyn Sunseri, and David~Ryan
  Koes.
\newblock Protein–ligand scoring with convolutional neural networks.
\newblock \emph{J Chemical Information and Modeling}, 2017.

\bibitem[Rao et~al.(2019)Rao, Bhattacharya, Thomas, Duan, Chen, Canny, Abbeel,
  and Song]{rao2019tape}
Roshan Rao, Nicholas Bhattacharya, Neil Thomas, Yan Duan, Xi~Chen, John Canny,
  Pieter Abbeel, and Yun~S. Song.
\newblock Evaluating protein transfer learning with tape.
\newblock \emph{Advances in Neural Information Processing Systems (NeurIPS)},
  2019.

\bibitem[Røgen and Fain(2003)]{rogen2003gauss}
Peter Røgen and Boris Fain.
\newblock Automatic classification of protein structure by using gauss
  integrals.
\newblock \emph{Proceedings of the National Academy of Sciences}, 2003.

\bibitem[Rives et~al.(2021)Rives, Meier, Sercu, Goyal, Lin, Liu, Guo, Ott,
  Zitnick, Ma, and Fergus]{rives2021esm}
Alexander Rives, Joshua Meier, Tom Sercu, Siddharth Goyal, Zeming Lin, Jason
  Liu, Demi Guo, Myle Ott, C.~Lawrence Zitnick, Jerry Ma, and Rob Fergus.
\newblock Biological structure and function emerge from scaling unsupervised
  learning to 250 million protein sequences.
\newblock \emph{Proceedings of the National Academy of Sciences}, 2021.

\bibitem[Russakovsky et~al.(2015)Russakovsky, Deng, Su, Krause, Satheesh, Ma,
  Huang, Karpathy, Khosla, Bernstein, Berg, and Fei-Fei]{ILSVRC15}
Olga Russakovsky, Jia Deng, Hao Su, Jonathan Krause, Sanjeev Satheesh, Sean Ma,
  Zhiheng Huang, Andrej Karpathy, Aditya Khosla, Michael Bernstein,
  Alexander~C. Berg, and Li~Fei-Fei.
\newblock {ImageNet Large Scale Visual Recognition Challenge}.
\newblock \emph{International Journal of Computer Vision (IJCV)}, 2015.

\bibitem[Shanehsazzadeh et~al.(2020)Shanehsazzadeh, Belanger, and
  Dohan]{shanehsazzadeh2020}
Amir Shanehsazzadeh, David Belanger, and David Dohan.
\newblock Is transfer learning necessary for protein landscape prediction?,
  2020.

\bibitem[Somnath et~al.(2021)Somnath, Bunne, and Krause]{somnath2021multiscale}
Vignesh~Ram Somnath, Charlotte Bunne, and Andreas Krause.
\newblock Multi-scale representation learning on proteins.
\newblock In A.~Beygelzimer, Y.~Dauphin, P.~Liang, and J.~Wortman Vaughan,
  editors, \emph{Advances in Neural Information Processing Systems}, 2021.

\bibitem[Strodthoff et~al.(2020)Strodthoff, Wagner, Wenzel, and
  Samek]{strodthoff2020udsmprot}
Nils Strodthoff, Patrick Wagner, Markus Wenzel, and Wojciech Samek.
\newblock Udsmprot: universal deep sequence models for protein classification.
\newblock \emph{Bioinformatics}, 2020.

\bibitem[Strokach et~al.(2020)Strokach, Becerra, Corbi-Verge, Perez-Riba, and
  Kim]{strokach2020protdes}
Alexey Strokach, David Becerra, Carles Corbi-Verge, Albert Perez-Riba, and
  Philip~M. Kim.
\newblock Fast and flexible protein design using deep graph neural networks.
\newblock \emph{Cell Systems}, 2020.

\bibitem[Townshend et~al.(2019)Townshend, Bedi, Suriana, and
  Dror]{townshend2019endtoend}
Raphael Townshend, Rishi Bedi, Patricia Suriana, and Ron Dror.
\newblock End-to-end learning on {3D} protein structure for interface
  prediction.
\newblock In \emph{Advances in Neural Information Processing Systems
  (NeurIPS)}, 2019.

\bibitem[Townshend et~al.(2020)Townshend, V{\"{o}}gele, Suriana, Derry, Powers,
  Laloudakis, Balachandar, Anderson, Eismann, Kondor, Altman, and
  Dror]{townshend2022atom3d}
Raphael J.~L. Townshend, Martin V{\"{o}}gele, Patricia Suriana, Alexander
  Derry, Alexander Powers, Yianni Laloudakis, Sidhika Balachandar, Brandon~M.
  Anderson, Stephan Eismann, Risi Kondor, Russ~B. Altman, and Ron~O. Dror.
\newblock {ATOM3D:} tasks on molecules in three dimensions.
\newblock \emph{arxiv}, 2020.

\bibitem[Tschannen et~al.(2020)Tschannen, Djolonga, Rubenstein, Gelly, and
  Lucic]{tschannen2020contrast}
Michael Tschannen, Josip Djolonga, Paul~K. Rubenstein, Sylvain Gelly, and Mario
  Lucic.
\newblock On mutual information maximization for representation learning.
\newblock \emph{International Conference on Learning Representations}, 2020.

\bibitem[van~den Oord et~al.(2018)van~den Oord, Li, and
  Vinyals]{vandenoord2018replearn}
A{\"{a}}ron van~den Oord, Yazhe Li, and Oriol Vinyals.
\newblock Representation learning with contrastive predictive coding.
\newblock 2018.

\bibitem[Van~der Maaten and Hinton(2008)]{van2008visualizing}
Laurens Van~der Maaten and Geoffrey Hinton.
\newblock Visualizing data using t-sne.
\newblock \emph{Journal of machine learning research}, 2008.

\bibitem[Vaswani et~al.(2017)Vaswani, Shazeer, Parmar, Uszkoreit, Jones, Gomez,
  Kaiser, and Polosukhin]{vaswani2017transf}
Ashish Vaswani, Noam Shazeer, Niki Parmar, Jakob Uszkoreit, Llion Jones,
  Aidan~N Gomez, \L~ukasz Kaiser, and Illia Polosukhin.
\newblock Attention is all you need.
\newblock In \emph{Advances in Neural Information Processing Systems}, 2017.

\bibitem[Wang et~al.(2013)Wang, Ma, Peng, and Xu]{wang2013protein}
Sheng Wang, Jianzhu Ma, Jian Peng, and Jinbo Xu.
\newblock Protein structure alignment beyond spatial proximity.
\newblock \emph{Scientific reports}, 2013.

\bibitem[Wang et~al.(2022)Wang, Combs, Brand, Calvo, Xu, Price, Golovach,
  Salawu, Wise, Ponnapalli, and Clark]{wang2022lmgvp}
Zichen Wang, Steven~A. Combs, Ryan Brand, Miguel~Romero Calvo, Panpan Xu,
  George Price, Nataliya Golovach, Emmanuel~O. Salawu, Colby~J. Wise, Sri~Priya
  Ponnapalli, and Peter~M. Clark.
\newblock Lm-gvp: an extensible sequence and structure informed deep learning
  framework for protein property prediction.
\newblock \emph{Scientific Reports}, 2022.

\bibitem[Wu et~al.(2022)Wu, Zhang, Radev, Wang, Jin, Jiang, Niu, and
  Li]{Wu2022md}
Fang Wu, Qiang Zhang, Dragomir Radev, Yuyang Wang, Xurui Jin, Yinghui Jiang,
  Zhangming Niu, and Stan~Z. Li.
\newblock Pre-training of deep protein models with molecular dynamics
  simulations for drug binding, 2022.

\bibitem[Xia et~al.(2022)Xia, Feng, Xia, Pan, and Shen]{Xia2022contrs}
Chunqiu Xia, Shi-Hao Feng, Ying Xia, Xiaoyong Pan, and Hong-Bin Shen.
\newblock Fast protein structure comparison through effective representation
  learning with contrastive graph neural networks.
\newblock \emph{PLOS Computational Biology}, 2022.

\bibitem[Ye et~al.(2019)Ye, Zhang, Yuen, and Chang]{ye2019embedding}
Mang Ye, Xu~Zhang, Pong~C. Yuen, and Shih{-}Fu Chang.
\newblock Unsupervised embedding learning via invariant and spreading instance
  feature.
\newblock \emph{IEEE Conference on Computer Vision and Pattern Recognition
  (CVPR)}, 2019.

\bibitem[You et~al.(2020)You, Chen, Sui, Chen, Wang, and Shen]{You2020GraphCL}
Yuning You, Tianlong Chen, Yongduo Sui, Ting Chen, Zhangyang Wang, and Yang
  Shen.
\newblock Graph contrastive learning with augmentations.
\newblock In \emph{Advances in Neural Information Processing Systems}, 2020.

\bibitem[Zeiler and Fergus(2014)]{zeiler2014cnn}
Matthew~D. Zeiler and Rob Fergus.
\newblock Visualizing and understanding convolutional networks.
\newblock In \emph{European Conference on Computer Vision (ECCV)}, 2014.

\bibitem[Zhang and Skolnick(2005)]{zhang2005tm}
Yang Zhang and Jeffrey Skolnick.
\newblock Tm-align: a protein structure alignment algorithm based on the
  tm-score.
\newblock \emph{Nucleic acids research}, 2005.

\bibitem[Zhang et~al.(2022)Zhang, Xu, Jamasb, Chenthamarakshan, Lozano, Das,
  and Tang]{Zhang2022}
Zuobai Zhang, Minghao Xu, Arian Jamasb, Vijil Chenthamarakshan, Aurelie Lozano,
  Payel Das, and Jian Tang.
\newblock Protein representation learning by geometric structure pretraining,
  2022.

\bibitem[Zhou et~al.(2019)Zhou, Barnes, Jingwan, Jimei, and
  Hao]{Zhou_2019_CVPR}
Yi~Zhou, Connelly Barnes, Lu~Jingwan, Yang Jimei, and Li~Hao.
\newblock On the continuity of rotation representations in neural networks.
\newblock In \emph{The IEEE Conference on Computer Vision and Pattern
  Recognition (CVPR)}, 2019.

\bibitem[Zotenko et~al.(2006)Zotenko, O'Leary, and
  Przytycka]{zotenko2006secstruc}
Elena Zotenko, Dianne~P. O'Leary, and Teresa~M. Przytycka.
\newblock Secondary structure spatial conformation footprint: a novel method
  for fast protein structure comparison and classification.
\newblock \emph{BMC Structural Biology}, 2006.

\bibitem[Öztürk et~al.(2018)Öztürk, Özgür, and
  Ozkirimli]{ozturk2018deepdta}
Hakime Öztürk, Arzucan Özgür, and Elif Ozkirimli.
\newblock {DeepDTA: deep drug–target binding affinity prediction}.
\newblock \emph{Bioinformatics}, 2018.

\end{thebibliography}
\bibliographystyle{plainnat}

\newpage

\appendix

\section{Network architecture} 
\label{sup:sec:network_arch}

In this section we describe the network architectures used in the experiments. All layers in our networks are followed by a batch normalization layer and a Leaky-ReLU activation function.

\begin{figure}[t]
  \begin{center}
    \includegraphics[width=\textwidth]{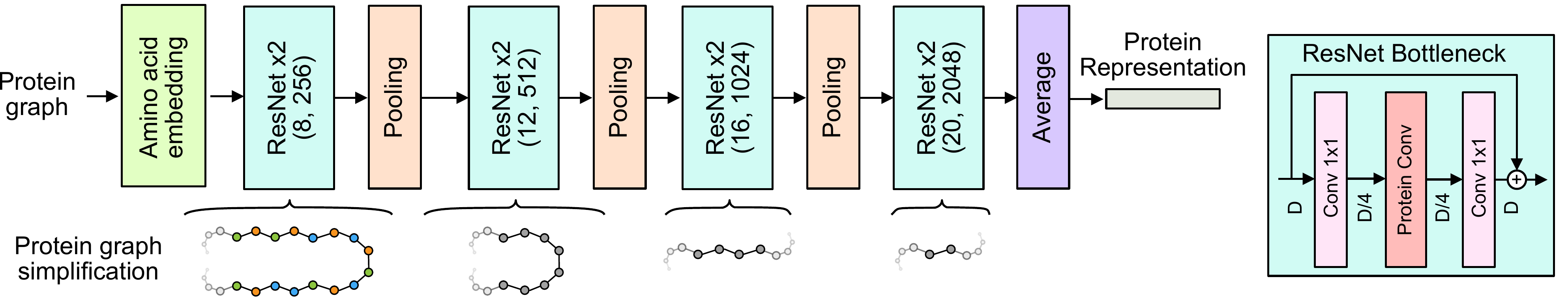}
      \caption{Illustration of our protein encoder. We use an amino acid embedding as our input features that are then processed by consecutive ResNet Bottleneck blocks and pooling operations. To obtain the final protein representation we use the average of the features from the remaining graph nodes.}
      \label{img:arch}
  \end{center}
\end{figure}

\paragraph{Protein encoder.}Our neural network receives as input the list of amino acids of the protein.
Each protein is then simplified several times with a pooling operation that reduces the number of amino acids by half each step.
We use the same pooling operation proposed by \citet{hermosilla2021ieconv} where every two consecutive amino acids are grouped into a new node.
The initial features are defined by an embedding of $16$ features for each amino acid type that is optimized together with the network parameters.
These initial features are then processed by two ResNet bottleneck blocks~\citep{he2016resnet} and then pooled to the next simplified protein representation using average pooling.
This process is repeated four times until we obtain a set of features for the last simplified protein graph.
The number of features used for each level are [$256, 512, 1024, 2048$].
The radius of the receptive field, $d$, used to compute the adjacency matrix $\mathcal B$ in each level are [$8, 12, 16, 20$] \angstrom.
Lastly, in order to obtain a set of features for the complete protein structure we use an order invariant operation that aggregates the features of all nodes.
In particular, we use the average of the features of all nodes.
Figure~\ref{img:arch} provides an illustration of the proposed architecture.
This model contains $30\,M$ parameters.

\paragraph{Protein encoder (reduced).}For the task of protein-ligand binding affinity, we use a reduced version of our protein encoder to avoid over-fitting due to the reduced number of proteins in the training set.
We use three pooling operations instead of four.
Moreover, we use one ResNet block per level instead of two ResNet bottleneck blocks in each level.
We use [$8, 12, 16$] \angstrom as distances $d$ in each level, and reduce the number of features to [$64, 128, 256$].
This results in a protein representation of $256$ features instead of $2048$ used for the other tasks.
This model contains $20\,M$ parameters.

\paragraph{Ligand encoder.}In the task of protein-ligand binding affinity, besides encoding the protein, we also have to encode the ligand.
As ligand encoder, we use a simple graph convolutional neural network architecture, with layers implemented as described in \citet{kipf2016semi}.
We use three layers with $64$, $64$, and $128$ output features.
The features of all layers are concatenated and the maximum and average are computed to obtain the final representation.
This model contains $15.7\,K$ parameters.

\paragraph{MLP.}For all tasks, the representation learned by the different encoders is further processed by a multi-layer perceptron (MLP) with one hidden layer.
The size of this hidden layer is defined as $(L_{in}*L_{out})/2$, where $L_{in}$ is the size of the input representation and $L_{out}$ the number of outputs of the MLP.
The number of parameters of this model varies depending on the number of outputs.

\section{Detailed experiments}

In this section, we describe in detail the experiments presented in the paper.

\subsection{Protein encoder pre-training.}

\paragraph{Data set.}Our main data set used for unsupervised learning is based on the PDB~\cite{berman2000pdb}. 
We collected $476,362$ different protein chains, each composed of at least $25$ of amino acids. 
This set of protein chains was later reduced for each task to avoid similarities with the different test sets, removing chains from the pre-training set based on the available annotations. 

\paragraph{Training.}To train our models with the contrastive learning objective we used a latent representation $h$ of size $2048$ and a projected representation $z$ of size $128$.
We use Stochastic Gradient Descent (SGD) optimizer with an initial learning rate of $0.3$ which was decreased linearly until $0.0001$ after a fourth of the total number of training steps.
We use a batch size of $256$ and a dropout rate of $0.2$ for the whole architecture.
Moreover, we used a weight decay factor of $1e-5$.
All networks were trained for $550\,K$ training steps, resulting in $6$ days of training.

\subsection{Ligand encoder pre-training.}

\paragraph{Data set.}For pre-training the ligand encoder, we used the \textit{in-vitro} subset of the ZINC20 database~\cite{irwin2020zinc20}, which contains $307,853$ molecules reported or inferred active in direct binding assays.

\paragraph{Training.}To perform the data transformations during contrastive learning, we remove atoms from the molecules with a probability $p$ randomly selected between $15\,\%$ and $0.30\,\%$.
Note that this approach is different than the one used for proteins, since these molecules are not formed by a chain of atoms and the number of atoms is significantly smaller than the number of nodes in the proteins.

Moreover, we used a latent representation $h$ of size $512$ and a projected representation $z$ of size $128$.
We use Stochastic Gradient Descent (SGD) optimizer with an initial learning rate of $0.3$ which was decreased linearly until $0.0001$ after a fourth of the total number of training steps.
We use a batch size of $512$ and a dropout rate of $0.2$ for the whole architecture.
Moreover, we used a weight decay factor of $1e-5$.
The network was trained for $550\,K$ training steps, resulting in $4$ hours of training.

\subsection{Protein structural similarity, DaliLite~\cite{holm2019dalilite}}

\paragraph{Data set.}This data set is composed of $140$ protein domains from the SCOPe database~\cite{murzin1995scope} for which similar proteins have to be found from a set of $15,211$ protein chains.
Moreover, they provide another set composed of $176,022$ protein chains that we use to train our distance metric.
In this benchmark, different similarity levels are considered based on the SCOPe classification hierarchy, Fold, Superfamily, and Family. 

\paragraph{Metric.}
To evaluate the performance of different methods on the DaliLite benchmark we use $F_{max}$ as defined by \citet{holm2019dalilite}.
We sort the $15\,$K proteins based on our distance metric to our target and use the following definition of $F_{max}$:

\begin{align}
    F_{max} &= \max_{n} \frac{2 p(n) r(n)}{ p(n) + r(n)} \nonumber \\
            &=  \max_{n} \frac{2 TP(n)}{ n + T} \nonumber
\end{align}

\noindent where $n$ is the rank of the query in the ordered list, i. e. the index of the protein in the sorted list.
For the $n$ first results in the ordered list, we define $p(n)$ as the precision, $r(n)$ as the recall, $TP(n)$ as the number of true positives pairs, and $T$ is the number of structures in the class.
We compute the final value for the test set by averaging the $F_{max}$ among the $140$ test protein domains.
For more details on this metric, we refer the reader to \citet{holm2019dalilite}. 

\paragraph{Pre-training data set.}For pre-training, we remove all proteins from the PDB set that are annotated with the same Fold as the $140$ protein domains. 
This resulted in $432,884$ protein chains.

\paragraph{Training.}We train the protein encoder described in Sect.~\ref{sup:sec:network_arch} followed by an MLP to generate $128$ features that we use to compute the cosine distance.
We train the model for $450$ epochs using Stochastic Gradient Descent (SGD) with an initial learning rate of $0.001$, which is decreased to $0.0001$ after $300$ epochs and decreased again to $0.00001$ after $400$ epochs.
To regularize the model, we use a dropout of $0.2$ and weight decay of $5e-4$.
Moreover, we use gradient clipping with a value of $10.0$.
In the fine-tuning setup, we use a warm-up stage of $25$ epochs in which we increase linearly the learning rate from $0.0$ to $0.001$, and we fix the mean and stddev of the batch normalization layers in the protein encoder.

\begin{wraptable}{r}{6cm}%
\vspace{-0,25cm}
\setlength{\tabcolsep}{3pt}
\caption{Comparison of sampling method used in the DaliLite data set~\cite{holm2019dalilite}.}%
\begin{center}
\begin{tabular}{lrrr}
    \toprule
    &
    \multicolumn{1}{c}{Fold} & 
    \multicolumn{1}{c}{Super.} & 
    \multicolumn{1}{c}{Fam.}
    \\
    \midrule
    Reg. sampling & 
    0.35 & 0.55 & 0.63\\
    Hier. sampling & 
    \textbf{0.66} & \textbf{0.70} & \textbf{0.75}\\
    \bottomrule
\end{tabular}%
\label{supp:tbl:samp_dalilite}%
\end{center}
\vspace{-0,25cm}
\end{wraptable}

During training, we use a batch size of $128$, which is composed of proteins from different folds.
In particular, we select $16$ folds in each batch and sample $8$ different proteins from it at different similarity levels.
We first select two different superfamilies from the fold, then we select two different families from each superfamily, and lastly, we select two proteins from each family.

\paragraph{Results.}Results of other methods reported in the paper are directly obtained from the benchmark~\cite{holm2019dalilite}.

\paragraph{Additional results.}We compare the performance of our hierarchical sampling method to simply selecting two random proteins from each fold.
We can see in Tbl.~\ref{supp:tbl:samp_dalilite} that our proposed sampling strategy vastly improves the performance of the models.

\subsection{Protein structural similarity, GraSR~\cite{Xia2022contrs}}

\paragraph{Data set.}This data set contains two sets, $13,265$ protein domains from the SCOPe database~\cite{murzin1995scope} used as queries for comparison and $1,914$ protein chains from the PDB database used as targets. 
Following \citet{Xia2022contrs}, we train our model on the $13,265$ protein domains.

\paragraph{Metric.} This data set considers a hit when the TMScore~\cite{zhang2005tm} between the target protein chain and the protein domain is higher than $0.9*m$, where $m$ is the maximum TMScore between the target protein and all the query protein domains in the main set.
As in \citet{Xia2022contrs}, performance is measured with the hit ratio of the first $k$ query proteins in the sorted list:

\begin{equation}
    HR_k = \frac{1}{T} \sum_{i=1}^{T} \frac{N_i^{'}(k)}{min(k, N_i)} \nonumber
\end{equation}

\noindent where $T$ is the number of target proteins in the test set, $N_i$ the total number of hits for the target protein $i$, and $N_i^{'}(k)$ the number of predicted hits for the $k$ first queries in the sorted query list.

\paragraph{Pre-training data set.}For pre-training, we remove all proteins from the PDB set that are annotated with the same Fold as the $1,914$ protein chains. 
This resulted in $395,534$ protein chains.

\paragraph{Training.}We train the protein encoder described in Sect.~\ref{sup:sec:network_arch} followed by two MLPs (one for the target and another one for the query) to generate $128$ features that we use to compute the cosine distance.
During training, we maximize the cosine distance if the query protein is annotated as a hit for the given target and minimize it otherwise.
We train the model for $350$ epochs and a batch size of $64$ using SGD with an initial learning rate of $0.3$, which is decreased to $0.03$ after $250$ epochs.
To regularize the model, we use a dropout of $0.2$ and weight decay of $1e-5$.
Moreover, we use gradient clipping with a value of $10.0$.
In the fine-tuning setup, we use a warm-up stage of $25$ epochs in which we increase linearly the learning rate from $0.0$ to $0.005$, and we fix the mean and stddev of the batch normalization layers in the protein encoder.

\paragraph{Results.}Results of other methods reported in the paper are directly obtained from \citet{Xia2022contrs}.

\subsection{Fold classification~\cite{hou2018deepsf}}

\paragraph{Data set.}This data set contains $16,712$ proteins domains of $1,195$ different folds from the SCOPe 1.75 database~\citep{murzin1995scope}. 
The data set provides a train, validation, and three different test sets with increasing difficulty based on the similarity between the proteins in the train and test sets: Protein, Family, and Superfamily.
The train set is composed of $12,312$ proteins whilst the validation set contains $736$ proteins.
The Superfamily test set contains $718$ proteins belonging to different superfamilies as the proteins in the train set, the Family test set contains $1,254$ proteins belonging to different families as the proteins in the train set, and the Protein test set contains $1,272$ from the same families as the proteins in the train set.

\paragraph{Metric.}Performance is measured with mean accuracy on the test sets. 

\paragraph{Pre-training data set.}For pre-training, we filtered the PDB data set and removed all annotated protein chains with the same folds as the proteins in the test sets. 
This procedure generated one PDB data set for each test set. 
The resulting data sets contain $377,271$ chains for the Superfamily test set, $313,616$ chains for the Family test set, and $324,304$ chains for the Protein test set.

\paragraph{Training.}We train the protein encoder described in Sect.~\ref{sup:sec:network_arch} followed by an MLP to predict the final classification probability.
We train the model for $400$ epochs and a batch size of $8$ using SGD with an initial learning rate of $0.001$ that we decrease to $0.0001$ after $100$ epochs and again to $0.00001$ after $300$ epochs.
To regularize the model, we use a dropout probability of $0.2$ in the protein encoder and $0.5$ in the MLP.
Moreover, we use a weight decay factor of $5e-4$ and gradient clipping with a value of $10.0$.
In the fine-tuning setup, we use a warm-up stage of $25$ epochs in which we increase linearly the learning rate from $0.0$ to $0.0005$.
This learning rate is then reduced to $0.00005$ after $300$ epochs.
As before, we fix the mean and stddev of the batch normalization layers in the protein encoder.

\paragraph{Results.}Results of other methods reported in the paper are directly obtained from \citet{hermosilla2021ieconv}. 
For completeness, we have included additional methods in Tbl.~\ref{supp:tbl:class_acc}.

\begin{table}%
\caption{Comparison of our network to other methods on the two classification tasks (protein fold and enzyme catalytic reaction classification) measured as mean accuracy.
}%
\begin{center}
\begin{tabular}{llrrrrr}
    \toprule
    & &
    \# params &
    \multicolumn{3}{c}{Fold}&
    \multicolumn{1}{c}{React.}\\
    \cmidrule(lr){4-6}
    & & & 
    \multicolumn{1}{c}{Fold} & 
    \multicolumn{1}{c}{Super.} & 
    \multicolumn{1}{c}{Fam.} & 
    \\
    \midrule
    & HHSuite & & 17.5\,\% & 69.2\,\% & 98.6\,\% & 82.6\,\%\\
    & TMalign & & 34.0\,\% & 65.7\,\% & 97.5\,\% & \\
    \midrule
    \multirow{9}{*}{Sequence} & 1DCNN~\cite{hou2018deepsf}&
    1.0\,M&
    40.9\,\%&
    50.7\,\%&
    76.2\,\%\\
    & 1DResNet~\cite{rao2019tape}$^*$ & 41.7\,M&
    17.0\,\% & 31.0\,\% & 77.0\,\% & 70.9\,\%\\
    & LSTM~\cite{rao2019tape}$^*$ & 43.0\,M&
    26.0\,\% & 43.0\,\% & 92.0\,\% & 79.9\,\%\\
    & Transformer~\cite{rao2019tape}$^*$ & 38.4\,M&
    21.0\,\% & 34.0\,\% & 88.0\,\% & 69.8\,\%\\
    & LSTM~\cite{bepler2019embedstruct}$^*$ & 31.7\,M&
    17.0\,\% & 20.0\,\% & 79.0\,\% & 74.3\,\%\\
    & LSTM~\cite{bepler2019embedstruct}$^{\dag}$& 31.7\,M&
    36.6\,\% & 62.7\,\% & 95.2\,\% & 66.7\,\%\\
    & mLSTM~\cite{alley2019unirep}$^*$ & 18.2\,M&
    23.0\,\% & 38.0\,\% & 87.0\,\% & 72.9\,\%\\
    & LSTM~\cite{strodthoff2020udsmprot}$^*$ & 22.7\,M&
    14.9\,\% & 21.5\,\% & 83.6\,\% & 73.9\,\%\\
    & Transformer~\cite{elnaggar2020prottrans}$^*$ & 420.0\,M&
    26.6\,\% & 55.8\,\% & 97.6\,\% & 72.2\,\%\\
    \cmidrule{1-7}
    \multirow{3}{*}{Structure} & GCNN~\cite{kipf2016semi} & 1.0\,M &
    16.8\,\% & 21.3\,\% & 82.8\,\% & 67.3\,\% \\
    & GCNN~\cite{diehl2019edge} & 1.0\,M &
    12.9\,\% & 16.3\,\% & 72.5\,\% & 57.9\,\% \\
    & 3DCNN~\cite{derevyanko2018prot3dcnn}  & 6.0\,M&
    31.6\,\% & 45.4\,\% & 92.5\,\% & 78.8\,\%\\
    \cmidrule{1-7}
    \multirow{6}{*}{Seq. + Struct.} & LSTM+GCNN~\cite{gligorijevic2021function}$^*$  & 6.2\,M&
    15.3\,\% & 20.6\,\% & 73.2\,\% & 63.3\,\%\\
    & GCNN~\cite{baldassarre2020graphqa} & 1.3\,M&
    23.7\,\% & 32.5\,\% & 84.4\,\% & 60.8\,\%\\
    & IEConv~\cite{hermosilla2021ieconv}  & 9.8\,M&
    45.0\,\% & 69.7\,\% & 98.9\,\% & 87.2\,\%\\
    \cmidrule{2-7}
    & Ours (\textit{no pre-train}) & 36.6\,M &
    \underline{47.6\,\%} & \underline{70.2\,\%} & \underline{99.2\,\%} & \underline{87.2\,\%}\\
    & Ours (\textit{MLP}) & 36.6\,M & 38.6\,\% & 69.3\,\% & 98.4\,\% & 80.2\,\%\\
    & Ours (\textit{fine-tune}) & 36.6\,M & \textbf{50.3\,\%} &  \textbf{80.6\,\%} & \textbf{99.7\,\%} & \textbf{88.1\,\%}\\
    \bottomrule
    \bottomrule
    \multicolumn{5}{l}{{$^*$}{\footnotesize Pre-trained unsupervised on 10-31 million protein sequences.}}\\
    \multicolumn{5}{l}{{$^\dag$}{\footnotesize Pre-trained on several supervised tasks with structural information.}}
\end{tabular}%
\label{supp:tbl:class_acc}%
\end{center}
\end{table}%

\subsection{Enzyme reaction classification~\cite{hermosilla2021ieconv}}

\paragraph{Data set.}This data set contains $37,428$ proteins from $384$ different EC numbers.
The task consists of, given a 3D structure of an enzyme, to predict its complete EC number, e.g. 4.2.3.1, among the $384$ numbers available in the data set.
The proteins in the data set are split into three sets, training, validation, and testing, whereby proteins in each set do not have more than $50\%$ of sequence similarity with proteins from the other sets. 
Thus, we obtain $29,215$ proteins for training, $2,562$ proteins for validation, and $5,651$ for testing. 

\paragraph{Metric.}Performance is measured with mean accuracy on the test set.

\paragraph{Pre-training data set.}For pre-training, we remove all proteins from the PDB data set which belong to the same EC number as the $384$ used in the test set, resulting in $359,909$ protein chains for pre-training.

\paragraph{Training.}For training this model, we use the same configuration and parameters as in the fold classification task.

\paragraph{Results.}Results of other methods reported in the paper are directly obtained from \citet{hermosilla2021ieconv}. 
For completeness, we also have included additional methods in Tbl.~\ref{supp:tbl:class_acc}.

\subsection{GO term prediction~\cite{gligorijevic2021function}}

\paragraph{Data set.}This data set contains $36,641$ proteins annotated with different Gene Ontology (GO) terms.
Proteins are selected to obtain between $50$ and $5,000$ proteins annotated for each individual GO term during training.
The GO is organized hierarchically, and the data set considers three different sub-branches on the hierarchy: Molecular Function (MF), Biological Process (BP), and Cellular Component (CC) terms.
Each sub-branch is predicted separately, using a different model to predict each term type separately.
The number of GO terms for each sub-branch is $489$ terms for the MF sub-branch, $1,943$ terms for the BP sub-branch, and $320$ terms for the CC sub-branch.
The data set is organized in training, validation, and test sets, each containing $29,902$, $3,323$, and $3,416$ proteins.
The proteins in the test set contain only experimentally determined functions and are classified based on their sequence similarity to the train set, having $<30\,\%$, $<40\,\%$, $<50\,\%$, $<70\,\%$, or $<95\,\%$ similarity.
Results are reported on the complete test set.
However, this classification allow us to evaluate the generalization ability of our pre-trained models for different sequence similarity levels.

\paragraph{Metric.}Performance is measured using protein-centric maximum F-Score, $F_{max}$, as defined in \citet{gligorijevic2021function}.
This metric measures the precision and recall of the predictions for each protein independently, and computes their mean.
These means are then used to compute the F-score.
$F_{max}$ is the maximum F-score among the different thresholds $t$ tested in the range $[0, 1]$.
Formally:

\begin{align}
    F_{max} &= \max_{t}\left\{ \frac{2 p(t) r(t)}{p(t) + r(t)} \right\} \nonumber \\
    p(t) &= \frac{1}{\vert M(t) \vert}\sum_{i \in M(t)}\frac{\vert TP_i(t) \vert}{\vert Pred_i(t) \vert} \nonumber \\
    r(t) &= \frac{1}{\vert N \vert}\sum_{i \in N}\frac{\vert TP_i(t) \vert}{\vert P_i \vert} \nonumber 
\end{align}

\noindent where $TP_i(t)$ if the set of correctly predicted GO terms for protein $i$ using threshold $t$, $Pred_i(t)$ if the set of predicted GO terms for protein $i$ using threshold $t$, $N$ is the set of all proteins in the test set, and $M(t)$ is the set of proteins for which we predict at least one GO term using threshold $t$.

\paragraph{Pre-training data set.}For pre-training, we remove all proteins from the PDB data set which 
contain the same annotations as the ones used in the test set.
The annotations, as in \citet{gligorijevic2021function}, are obtained from the SIFTS database~\cite{Dana2019SIFTS}.
This filtering results in $391,882$ protein chains for pre-training.

\paragraph{Training.}We train the protein encoder described in Sect.~\ref{sup:sec:network_arch} followed by an MLP to predict the probability of each of the different GO terms.
We train the model for $900$ epochs and a batch size of $64$ using SGD with an initial learning rate of $0.001$ that we decrease to $0.0001$ after $700$ epochs.
To regularize the model, we use a dropout probability of $0.2$ in the protein encoder and $0.3$ in the MLP.
Moreover, we use a weight decay factor of $5e-4$ and gradient clipping with a value of $10.0$.
In the fine-tuning setup, as in the other tasks, we use a warm-up stage of $25$ epochs in which we increase linearly the learning rate from $0.0$ to the initial learning rate.
For the MF and CC data sets we use an initial learning rate of $0.0005$ which is reduced to $0.00005$ after $500$ epochs.
For the BB data set we use the same initial learning rate as when we train the model from scratch.
As in the other tasks, we fix the mean and stddev of the batch normalization layers in the protein encoder.

\paragraph{Results.}Performance of other methods are obtained from the works of \citet{wang2022lmgvp} and \citet{Zhang2022}.
For completeness, we include additional comparisons in Tbl.~\ref{supp:tbl:go_terms}.

\begin{table}
\caption{$F_{max}$ of our method on the GO term prediction tasks~\cite{gligorijevic2021function} compared to other methods, pre-trained or not.}%
\begin{center}
\begin{tabular}{llrrr}
    \toprule
    & & 
    \multicolumn{1}{c}{MF} & 
    \multicolumn{1}{c}{BP} & 
    \multicolumn{1}{c}{CC}\\
    \midrule
    \multicolumn{5}{c}{No pre-training}\\
    \midrule
    \multirow{4}{*}{Sequence-based} & CNN~\cite{shanehsazzadeh2020} & 0.354 & 0.244 & 0.387 \\
    & ResNet~\cite{rao2019tape} & 0.267 & 0.280 & 0.403 \\
    & LSTM~\cite{rao2019tape} & 0.166 & 0.248 & 0.320 \\
    & Transformer~\cite{rao2019tape} & 0.240 & 0.257 & 0.380 \\
    \midrule
    \multirow{5}{*}{Structure-based} & GAT~\cite{Zhang2022} & 0.317 & 0.284 & 0.385 \\
    & GVP~\cite{Zhang2022} & 0.426 & 0.326 & 0.420 \\
    & DeepFriGO~\cite{Zhang2022} & 0.465 & 0.399 & \textbf{0.460} \\
    & GearNet~\cite{Zhang2022} & \underline{0.580} & \underline{0.403} & \underline{0.450} \\
    \cmidrule{2-5}
    & Ours & \textbf{0.624} & \textbf{0.421} & 0.431 \\
    \midrule
    \multicolumn{5}{c}{With pre-training}\\
    \midrule
    \multirow{3}{*}{Sequence pre-train} & ESM-1b~\cite{rives2021esm} & \underline{0.657} & \underline{0.470} & 0.488 \\
    & LM-GVP~\cite{wang2022lmgvp} & 0.545 & 0.417 & \textbf{0.527} \\
    & ProtBERT-BFD~\cite{elnaggar2020prottrans} & 0.456 & 0.279 & 0.408 \\
    \midrule
    \multirow{3}{*}{Structure pre-train} & GearNet~\cite{Zhang2022} & 0.650 & \textbf{0.490} & 0.486 \\
    \cmidrule{2-5}
    & Ours (\textit{MLP}) & 0.606 & 0.443 & 0.506 \\
    & Ours (\textit{fine-tune}) & \textbf{0.661} & 0.468 & \underline{0.516}\\
    \bottomrule
\end{tabular}%
\label{supp:tbl:go_terms}%
\end{center}
\end{table}%

\subsection{Protein--Ligand binding affinity prediction~\cite{townshend2022atom3d}}

\paragraph{Data set.}This data set contains $4,709$ filtered pairs of protein-ligand from the PDBind database (version 2019)~\cite{liu2017pdbbind} with their corresponding binding affinity, expressed in molar units of the inhibition constant ($K_i$) or dissociation constant ($K_d$).
As in previous work~\cite{somnath2021multiscale,townshend2022atom3d,Wu2022md}, we do not distinguish between these constants and predict the negative log transformation of these, $pK =-log(K)$.
This task contains two different data sets with different maximum sequence similarity between the train and test set, $60\%$ and $30\%$.
Each of these is further divided into train, validation, and test sets, each containing $3,507$, $466$, and $490$ respectively for the $30\,\%$ data set, and $3,678$, $460$, and $460$ for the $60\,\%$ data set.

\paragraph{Metric.}Performance is measured with root mean squared error (RMSE), and Pearson and Spearman correlation coefficients.

\paragraph{Pre-training data set.}Since the annotations on this data set are not categorical variables, we filtered the PDB data set used for pre-training based on sequence similarity.
We remove all protein chains from our pre-training data set that have a sequence similarity higher than $30\,\%$ to any protein in the two test sets.
This resulted in $385,592$ protein chains used for pre-training the model.

\paragraph{Training.}We train the protein and ligand encoders described in Sect.~\ref{sup:sec:network_arch} followed by an MLP to create the protein and ligand representations.
We define the final binding affinity as the dot product between these representations, re-scaled to the range defined by the maximum and minimum binding affinity in the training data set.
We train the model for $250$ epochs and a batch size of $32$ using Adam optimizer with an initial learning rate of $0.001$.
We apply a learning rate decay of $0.9$ based on a validation Spearman correlation plateau, with a patience of $5$ epochs.
To regularize the model, we use a dropout probability of $0.2$, a weight decay factor of $1e-5$, and gradient clipping with a value of $10.0$.
Moreover, following \citet{somnath2021multiscale}, we apply noise to the gradients with a decreasing factor wrt. the training epoch.
In the fine-tuning setup, as in the other tasks, we use a warm-up stage of $15$ epochs in which we increase linearly the learning rate from $0.0$ to $0.0001$.
As in the other tasks, we fix the mean and stddev of the batch normalization layers in the protein encoder.

\paragraph{Results.}Performance of other methods are obtained from the works of \citet{somnath2021multiscale} and \citet{Wu2022md}.
For completeness, we include additional comparisons in Tbl.~\ref{supp:tbl:prot_lig}.

\begin{table}%
\caption{RMSE, Pearson, and Spearman coefficients on the protein-ligand binding affinity prediction task~\cite{townshend2022atom3d}. Comparison to different 3D structure and sequence based methods, and a method pre-trained on molecular dynamics simulations (PretrainMD~\cite{Wu2022md}).}%
\setlength{\tabcolsep}{4.5pt}
\begin{center}
\footnotesize
\begin{tabular}{lrrrrrr}
    \toprule
    \toprule
    &
    \multicolumn{3}{c}{Seq. Id. ($60\,\%$)} &
    \multicolumn{3}{c}{Seq. Id. ($30\,\%$)}
    \\
    &
    \multicolumn{1}{c}{RMSE $\downarrow$} &
    \multicolumn{1}{c}{Pears. $\uparrow$} &
    \multicolumn{1}{c}{Spear. $\uparrow$} &
    \multicolumn{1}{c}{RMSE $\downarrow$} &
    \multicolumn{1}{c}{Pears. $\uparrow$} &
    \multicolumn{1}{c}{Spear. $\uparrow$}
    \\
    \midrule
    \multicolumn{7}{l}{Sequence-based} \\ 
    \midrule
    DeepDTA~\cite{ozturk2018deepdta} & 
    1.762 {$\pm$ \scriptsize .261} & 0.666 {$\pm$ \scriptsize .012} & 0.663 {$\pm$ \scriptsize .015} &
     1.565 {$\pm$ \scriptsize .080} & \underline{0.573 {$\pm$ \scriptsize .022}} & 0.574 {$\pm$ \scriptsize .024}\\
    LSTM~\cite{bepler2019embedstruct} & 
    1.891 {$\pm$ \scriptsize .004} & 0.249 {$\pm$ \scriptsize .006} & 0.275 {$\pm$ \scriptsize .008} &
     1.985 {$\pm$ \scriptsize .016} & 0.165 {$\pm$ \scriptsize .006} & 0.152 {$\pm$ \scriptsize .024}\\
    TAPE~\cite{rao2019tape} & 
    1.633 {$\pm$ \scriptsize .016} & 0.568 {$\pm$ \scriptsize .033} & 0.571 {$\pm$ \scriptsize .021} &
     1.890 {$\pm$ \scriptsize .035} & 0.338 {$\pm$ \scriptsize .044} & 0.286 {$\pm$ \scriptsize .124}\\
    ProtTrans~\cite{elnaggar2020prottrans} & 
    1.641 {$\pm$ \scriptsize .016} & 0.595 {$\pm$ \scriptsize .014} & 0.588 {$\pm$ \scriptsize .009} &
     1.544 {$\pm$ \scriptsize .015} & 0.438 {$\pm$ \scriptsize .053} & 0.434 {$\pm$ \scriptsize .058}\\
    \midrule
    \multicolumn{7}{l}{Structure-based} \\ 
    \midrule
    3DCNN~\cite{townshend2022atom3d} & 
    1.450 {$\pm$ \scriptsize .024} & 0.716 {$\pm$ \scriptsize .008} & 0.714 {$\pm$ \scriptsize .009} &
    \underline{1.429 {$\pm$ \scriptsize .042}} & 0.541 {$\pm$ \scriptsize .029} & 0.532 {$\pm$ \scriptsize .033}\\
    3DGCNN~\cite{townshend2022atom3d} & 
    1.493 {$\pm$ \scriptsize .010} & 0.669 {$\pm$ \scriptsize .013} & 0.691 {$\pm$ \scriptsize .010} &
    1.963 {$\pm$ \scriptsize .120} & \textbf{0.581 {$\pm$ \scriptsize .039}} & \textbf{0.647 {$\pm$ \scriptsize .071}}\\
    Masif~\cite{gainza2020masif} & 
    1.426 {$\pm$ \scriptsize .017} & 0.709 {$\pm$ \scriptsize .008} & 0.701 {$\pm$ \scriptsize .011} &
    1.484 {$\pm$ \scriptsize .006} & 0.467 {$\pm$ \scriptsize .020} & 0.455 {$\pm$ \scriptsize .014}\\
    HoloProt~\cite{somnath2021multiscale} & 
    1.365 {$\pm$ \scriptsize .038} & 0.749 {$\pm$ \scriptsize .014} & 0.742 {$\pm$ \scriptsize .011} &
    1.464 {$\pm$ \scriptsize .006} & 0.509 {$\pm$ \scriptsize .002} & 0.500 {$\pm$ \scriptsize .005}\\
    PretrainMD~\cite{Wu2022md} & 
    1.468 {$\pm$ \scriptsize .026} & 0.673 {$\pm$ \scriptsize .015} & 0.691 {$\pm$ \scriptsize .014} &
    \textbf{1.419 {$\pm$ \scriptsize .027}} & 0.551 {$\pm$ \scriptsize .045} & \underline{0.575 {$\pm$ \scriptsize .033}}\\
    \midrule
    Ours (\textit{no pre-train}) & 
    \underline{1.347 {$\pm$ \scriptsize .018}} & 0.757 {$\pm$ \scriptsize .005} & 0.747 {$\pm$ \scriptsize .004} &
    1.589 {$\pm$ \scriptsize .081} & 0.455 {$\pm$ \scriptsize .045} & 0.451 {$\pm$ \scriptsize .043}\\
    Ours (\textit{MLP}) & 
    1.361 {$\pm$ \scriptsize .032} & \underline{0.763 {$\pm$ \scriptsize .009}} & \underline{0.763 {$\pm$ \scriptsize .010}} &
    1.525 {$\pm$ \scriptsize .070} & 0.498 {$\pm$ \scriptsize .036} & 0.493 {$\pm$ \scriptsize .044}\\
    Ours (\textit{fine-tune}) & 
    \textbf{1.332 {$\pm$ \scriptsize .020}} & \textbf{0.768 {$\pm$ \scriptsize .006}} & \textbf{0.764 {$\pm$ \scriptsize .006}} &
    1.452 {$\pm$ \scriptsize .044} & 0.545 {$\pm$ \scriptsize .023} & 0.532 {$\pm$ \scriptsize .025}\\
    \bottomrule
\end{tabular}%
\label{supp:tbl:prot_lig}%
\end{center}
\end{table}

\section{Ablation studies} 

In this section, we evaluate the design decisions of our framework.
We perform all our ablation experiments by training an MLP to classify the protein representations according to the Fold classification task.
For pre-training, we use the entire PDB data set, and train the protein encoder for $180\,$K training steps, which results in two days of computation.

\paragraph{Data transformation in the contrastive learning setup.}
First, we analyze how the amount of information removed from the sequence affects the learned representation.
From the results in Table~\ref{sup:tbl:ablations_data}, we can see that removing between $20\%$ and $40\%$ of the protein chain makes the contrastive objective too easy, and the protein encoder does not learn a rich enough representation.
On the other hand, removing between $60\%$ and $80\%$ does not preserve enough information, and the performance also suffers.
As can be seen, we found that removing between $40\%$ and $60\%$ of the protein chains achieves the best performance.

\begin{table}%
\parbox{.45\linewidth}{
\setlength{\tabcolsep}{3pt}
\caption{Ablations on the data transformations used during pre-training for the Fold Classification task.}%
\begin{center}
\begin{tabular}{crrrr}
    \toprule
    \multicolumn{1}{c}{Transf.}&
    \multicolumn{1}{c}{Super.} & 
    \multicolumn{1}{c}{Fam.} & 
    \multicolumn{1}{c}{Protein} & 
    \multicolumn{1}{c}{Avg}
    \\
    \midrule
    60\%-80\% &
    35.2\,\% & 67.2\,\% & 98.0\,\% & 66.8\,\% \\
    40\%-60\% &
    38.9\,\% & \textbf{70.1}\,\% & \textbf{98.8}\,\% & \textbf{69.3}\,\%\\
    20\%-40\% &
    32.5\,\% & 57.0\,\% & 97.6\,\% & 62.4\,\%\\
    Sub-graph &
    \textbf{42.1}\,\% & 65.9\,\% & 98.7\,\% & 68.9\,\%\\
    \bottomrule
\end{tabular}%
\label{sup:tbl:ablations_data}%
\end{center}
}
\hfill
\parbox{.45\linewidth}{
\setlength{\tabcolsep}{3pt}
\caption{Ablations on the data transformations used on the supervised setting for the Fold Classification task.}%
\begin{center}
\begin{tabular}{crrrr}
    \toprule
    \multicolumn{1}{c}{Transf.}&
    \multicolumn{1}{c}{Super.} & 
    \multicolumn{1}{c}{Fam.} & 
    \multicolumn{1}{c}{Protein} & 
    \multicolumn{1}{c}{Avg}
    \\
    \midrule
    60\%-80\% &
    9.9\,\% & 23.8\,\% & 62.7\,\% & 32.1\,\% \\
    40\%-60\% &
    26.7\,\% & 48.3\,\% & 91.9\,\% & 55.6\,\% \\
    20\%-40\% &
    38.9\,\% & 63.2\,\% & 98.0\,\% & 66.7\,\% \\
    Noise &
    \textbf{47.6}\,\% & \textbf{70.2}\,\% & \textbf{99.2}\,\% & \textbf{72.3}\,\%\\
    \bottomrule
\end{tabular}%
\label{sup:tbl:ablations_data_sup}%
\end{center}
}
\end{table}%

We further compare our suggested data transformation approach to the graph augmentation technique used by \citet{You2020GraphCL}. 
\citet{You2020GraphCL} selected a random sub-graph on the spatial neighboring graph, thus selecting a random area in 3D space.
We can see in Table~\ref{sup:tbl:ablations_data}, that while \citet{You2020GraphCL} obtains a good performance on the Superfamily test set, it obtains lower accuracy on the Family and Protein test set.
Since the Family and Protein test sets contain proteins with higher sequence similarity to the training set than the Superfamily test set, we hypothesize that our method uses more information of the protein sequence for the representation than the sub-graph method, since the latest sees disconnected sections of the chains during training, while our method always sees a connected sub-chain.
Although we acknowledge that both methods can be beneficial for different tasks, we observed that on average our method provides better performance.

\paragraph{Data transformation in supervised setup.}
In this experiment, we evaluate how the proposed data transformation could affect the supervised training of the protein encoder on the Fold Classification task.
We can see in Table~\ref{sup:tbl:ablations_data_sup}, that in contrast to the unsupervised training, our data transformation technique used as data augmentation reduces performance in this setup. 
Instead, the best accuracy is obtained by adding a small random Gaussian noise into the 3D coordinates of the alpha carbon.
These results align with the ones obtained on other contrastive learning works~\citep{chen2020simple}, where these extreme data transformation strategies hurt the supervised training instead of improving its performance.

\paragraph{Edge features.}
In this ablation study, we evaluate the different components of our convolution operation (see Table~\ref{sup:tbl:ablations_conv}).
Our baseline method uses as edge features the Euclidean distance and the shortest path along the sequence as defined by \citet{hermosilla2021ieconv}.
We evaluate the performance improvement when we substitute the Euclidean distance by direction and orientation information as described in Sect. 3.4 in the main paper, denoted as \textit{Rot. Eq.} in the table.
We also evaluate how transforming the original inputs similar to positional encoding affects the resulting performance, denoted as \textit{Add. Input} in the table.
Furthermore, we evaluate the effect of the smoothing function applied towards the boundary of the receptive field, denoted as \textit{Smooth} in Table~\ref{sup:tbl:ablations_conv}.
We can see that adding the components individually to the baseline, results in an improvement of accuracy in all cases. 
Moreover, when we incorporate all together in our final convolution operation, we experience even a higher improvement, \textit{Full} in Table~\ref{sup:tbl:ablations_conv}.

\begin{table}%
\parbox{.45\linewidth}{
\setlength{\tabcolsep}{3pt}
\caption{Ablations on the edge feature elements on the Fold Classification task.}%
\begin{center}
\begin{tabular}{lrrrr}
    \toprule
    &
    \multicolumn{1}{c}{Super.} & 
    \multicolumn{1}{c}{Fam.} & 
    \multicolumn{1}{c}{Protein} & 
    \multicolumn{1}{c}{Avg}
    \\
    \midrule
    Baseline &
    39.0\,\% & 65.6\,\% & 98.6\,\% & 67.7\,\%\\
    Rot. Eq. & 
    45.5\,\% & 69.7\,\% & 98.9\,\% & 71.4\,\%\\
    Add. Input & 
    44.6\,\% & 67.7\,\% & 98.7\,\% & 70.3\,\% \\
    Smooth & 
    40.7\,\% & 65.4\,\% & 98.4\,\% & 68.2\,\% \\
    Full &
    \textbf{47.6}\,\% & \textbf{70.2}\,\% & \textbf{99.2}\,\% & \textbf{72.3}\,\% \\
    \bottomrule
\end{tabular}%
\label{sup:tbl:ablations_conv}%
\end{center}
}
\hfill
\parbox{.45\linewidth}{
\setlength{\tabcolsep}{3pt}
\caption{Ablations of the rotation representation used on the Fold Classification task.}%
\begin{center}
\begin{tabular}{crrrr}
    \toprule
     & 
    \multicolumn{1}{c}{Super.} & 
    \multicolumn{1}{c}{Fam.} & 
    \multicolumn{1}{c}{Protein} & 
    \multicolumn{1}{c}{Avg}
    \\
    \midrule
    Quat. & 
    44.0\,\% & 68.7\,\% & 98.0\,\% & 70.2\,\%\\
    6D & 
    \textbf{46.1}\,\% & 68.2\,\% & 98.7\,\% & 71.0\,\%\\
    Dot Axis & 
    45.5\,\% & \textbf{69.7}\,\% & \textbf{98.9}\,\% & \textbf{71.4}\,\%\\
    \bottomrule
\end{tabular}%
\label{sup:tbl:ablations_rot}%
\end{center}
}
\end{table}%

\paragraph{Edge orientation representation.}
Lastly, we evaluated the performance of the model, when changing the representation of the orientation features.
Here, we compare representing the orientation with Quaternions~\cite{Ingraham2019genprot}, with the 6D representation introduced by \citet{Zhou_2019_CVPR}, and the simple dot product between the axes of the two frames used in this paper.
Results of this experiment are shown in Table~\ref{sup:tbl:ablations_rot}.
The worse performance is obtained by Quaternions, while the 6D and the dot product obtain similar performance.
Although the dot product is not able to represent a full rotation, it obtained a slightly higher performance than the 6D and a faster convergence during training.
We hypothesize, that even the 6D representation is more descriptive, it uses more floats than the dot product method wrt. the rest of the inputs to the kernel.

\begin{figure}
  \begin{center}
    \includegraphics[width=\textwidth]{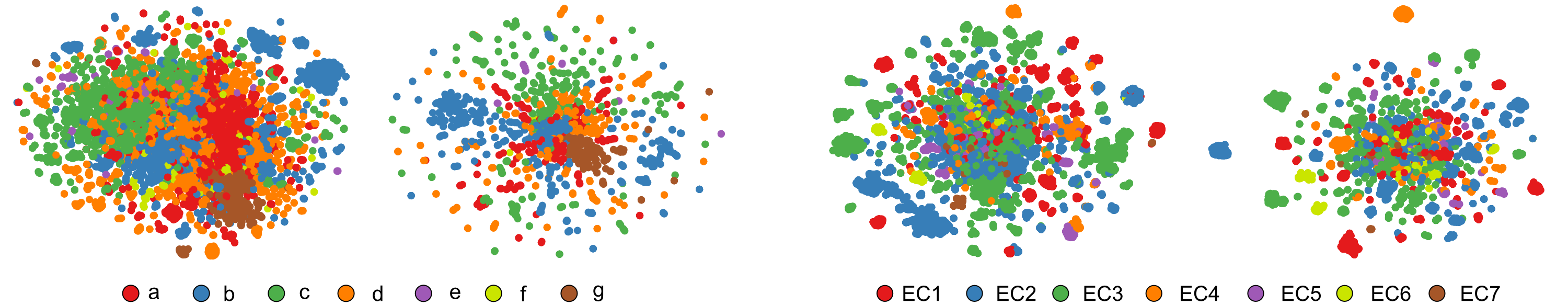}
      \caption{Dimensionality reduction of the protein representations using TSNE~\citep{van2008visualizing}. \textbf{Left:} Proteins from the Fold Classification task, training set on the left, test set on the right, color-coded based on the highest hierarchy level in the SCOPe classification system. \textbf{Right:} Proteins from the Enzyme Classification task, training on the left, test set on the right, color-coded based on the highest level of the Enzyme Commission number.}
      \label{img:tsne}
  \end{center}
\end{figure}

\section{Latent space visualization}
We visualize the representation learned by mapping the high dimensional space to a 2D representation using TSNE~\citep{van2008visualizing}. 
Then, we color-code each point based on the SCOPe and EC number classification schemes (see Figure~\ref{img:tsne}).

For the Fold Classification task, we take the training and test set Protein, and color-code each data point based on their class according to the SCOPe classification hierarchy. 
Note, that the model did not see during training any of the folds of the proteins in the test set. 
We can see, that our representation clusters points from the same class for classes \textit{a}, \textit{b}, \textit{c}, \textit{d}, and \textit{g}. 
However, points from classes \textit{e} and \textit{f} are spread among the other classes.

Moreover, we also use the same visualization for the Enzyme Classification task. 
We color code each data point based on the first number from the EC number. 
Figure~\ref{img:tsne} shows that, even if the data points do not seem to form a unique cluster for each EC number, data points from small clusters in the embedding all belong to the same EC class.
This might be an indication, that the network did not use the higher levels of the EC number classification scheme to cluster data points, but groups proteins based on other properties that are captured beyond the first digit of the EC number.

\end{document}